\documentclass[12pt]{JHEP3}
\usepackage{graphicx}
\usepackage{amsmath}

\normalbaselines

\title{Feynman Rules for the Rational Part of \\  the Standard Model One-loop
Amplitudes in the 't Hooft-Veltman  $\gamma_5$ Scheme}

\author{Hua-Sheng Shao\\
Department of Physics and State Key Laboratory of Nuclear Physics
and Technology, Peking University,
 Beijing 100871, China\\
E-mail: \email{erdissshaw@gmail.com}}

\author{Yu-Jie Zhang\\
Key Laboratory of Micro-nano
 Measurement-Manipulation and Physics (Ministry of Education) and School of Physics, Beihang University,
 Beijing 100191, China\\
E-mail: \email{nophy0@gmail.com}}

\author{Kuang-Ta Chao\\
 Department of Physics and State Key Laboratory of
Nuclear Physics and Technology, Peking University,
 Beijing 100871, China\\
Center for High Energy Physics, Peking University, Beijing 100871,
China\\
 E-mail: \email{ktchao@pku.edu.cn}}

\abstract{ We study Feynman rules for the rational part $R$ of the
Standard Model amplitudes at one-loop level in the 't Hooft-Veltman
$\gamma_5$ scheme. Comparing our results for quantum chromodynamics
and electroweak 1-loop amplitudes with that obtained based on the
Kreimer-Korner-Schilcher (KKS) $\gamma_5$ scheme, we find the latter
result can be recovered when our $\gamma_5$ scheme becomes identical
(by setting $g5s=1$ in our expressions) with the KKS scheme. As an
independent check, we also calculate Feynman rules obtained in the
KKS scheme, finding our results in complete agreement with formulae
presented in the literature. Our results, which are studied in two
different $\gamma_5$ schemes, may be useful for clarifying the
$\gamma_5$ problem in dimensional regularization. They are helpful
to eliminate or find  ambiguities arising from different dimensional
regularization schemes. }


\begin{document}
\section{Introduction}

The operation of the Large Hadron Collider (LHC) at CERN has opened
a new era in high energy physics. There are many important topics
for LHC physics, particularly in hunting Higgs boson(s) and
searching for new physics signals beyond the standard model (BSM).
However, the existence of large standard model (SM) background makes
such studies difficult. In this regard, the next-to-leading order
(NLO) corrections to the standard model calculations are necessary
and helpful, because these corrections may provide a much improved
scale dependence and more reliable estimations in separating SM
background from new physics signals.  Since the Les Houches workshop
in 2005, many NLO corrections in processes with four or more final
state particles at the LHC have been successfully performed. They
are, e.g., $pp\rightarrow t\bar{t}b\bar{b}+X$
\cite{Bredenstein:2008zb,Bevilacqua:2009zn}, $pp\rightarrow V+3jets$
\cite{KeithEllis:2009bu}, $pp\rightarrow W^{\pm}+4jets$
\cite{Berger:2010zx}, $pp\rightarrow VV+2jets$
\cite{Melia:2010bm,Campanario:2011cs}, $pp\rightarrow
t\bar{t}+2jets$ \cite{Bevilacqua:2010ve}, $pp(\bar{p})\rightarrow
W^+W^-b\bar{b}$ \cite{Denner:2010jp}, $pp\rightarrow
b\bar{b}b\bar{b}+X$ \cite{Binoth:2009rv} and $pp \rightarrow W
\gamma\gamma+jet$\cite{Campanario:2011ud}. However, a few SM
processes of similar complexity, which are relevant to search for
Higgs boson(s) or detection of new particles in physics BSM, are
still not available at NLO level \cite{Binoth:2010ra}, such as
$pp\rightarrow t\bar{t}t\bar{t}$.

Recently, automatic one-loop calculations have become a feasible
task after several new and efficient algorithms have been proposed.
Some of the most notable methods are the Unitarity
\cite{Bern:1993mq} based techniques such as the Ossola,
Papadopoulos, and Pittau (OPP) reduction method
\cite{delAguila:2004nf}, which reduces the computation of one-loop
amplitudes to a problem with a complexity similar to a tree level
calculation. Most of the aforementioned processes are computed
within this method or its variations.
 Such
approaches rely on the master integrals of one-loop amplitudes that
can be simply represented as a linear combination of up to 4-point
scalar integrals, i.e., those of boxes, triangles, bubbles and
tadpoles. A one-loop amplitude $\cal A$ can be written as
\begin{center}
\begin{eqnarray*}
\cal
A&=&\sum_{i}d_i~Box_i+\sum_{i}c_i~Triangle_i+\sum_{i}b_i~Bubble_i+\sum_{i}a_i~Tadpole_i+R.
\end{eqnarray*}
\end{center}
In the OPP framework, the coefficients $d_i$, $c_i$, $b_i$, $a_i$
and $R_1$ ($R=R_1+R_2$) can be obtained from the cut constructible
(CC) part of the amplitude, where the numerators of loop integrals
are dealt with in 4 dimensions, and $R_2$ should be derived
separately \cite{Ossola:2008xq}. With the ultraviolet nature of
$R_2$ \cite{Bredenstein:2008zb,Binoth:2006hk}, we can establish
Feynman rules for this part. Here we directly call $R_2$  the
rational term $R$ for brevity. In fact, this approach is more
efficient than applying the Unitarity reduction method in
$d=4-2\epsilon$ dimensions in practical calculations
\cite{Binoth:2010ra}. Algebraically, even in other methods, such as
the conventional Passarino-Veltman reduction procedure
\cite{Passarino:1978jh}, calculations of $R$ are also useful in
multi-leg processes \cite{Bredenstein:2008zb}.

Several previous works studied calculations of the Feynman rules for
$R$, and a set of Feynman rules for $R$ in the Standard Model (SM)
under one $\gamma_5$ strategy in the 't Hooft-Feynman gauge, $R_\xi$
gauge, and Unitary gauge has been reported
\cite{Draggiotis:2009yb,Garzelli:2009is,Garzelli:2010qm}, and a
package $R2SM$ written in FORM is also available
\cite{Garzelli:2010fq}. Moreover, some simplifications in extracting
rational terms were suggested recently in
Ref.\cite{Campanario:2011cs}. The primary aim of the present paper
is to give our result in another $\gamma_5$ regularization scheme,
i.e. the 't Hooft -Veltman $\gamma_5$ scheme. In addition, we would
also like to check the results presented in
Refs.\cite{Draggiotis:2009yb,Garzelli:2009is}.

This paper is organized as follows. In Section \ref{sec:2}, we
recall the theory for the rational part $R$, and set our
regularization schemes and $\gamma_5$ strategies. The notations and
Feynman rules are given in Section \ref{sec:3}. Our results in two
$\gamma_5$ schemes are expressed in Section \ref{sec:4}. Finally, a
summary is given in Section \ref{sec:5}.

\section{The theory of $R$ and regularization schemes \label{sec:2}}
In this section, we will describe the origin of the rational part
$R$, and discuss the dimensional regularization scheme (DREG) in two
$\gamma_5$ strategies. Due to the divergences in a one-loop
calculation, we compute our one-loop integrals in $d=4-2\epsilon$
dimensions to regularize the divergences that appear in 4
dimensions. A generic N-point one-loop (sub-) amplitude reads
\begin{eqnarray}
{\cal A}_N &=&{\int d^d \bar{q} \frac{\bar{N}(\bar{q})}{\bar{D}_1
\bar{D}_2 ...\bar{D}_N}},
\bar{D}_k=(\bar{q}+p_k)^2-(m_k)^2,\hspace{0.5 cm}
\bar{q}=q+\tilde{q},
\end{eqnarray}
where the bar means quantities defined in $d$ dimensions, while the
tilde in $d-4$ dimensions. All quantities of the external particles
such as external polarization vectors and external momenta $p_i$ are
maintained in 4 dimensions.

In numerator the function $\bar{N}(\bar{q})$ can be split into a
4-dimensional part $N(q)$ and a $(d-4)$-dimensional part
$\tilde{N}(\tilde{q}^2,\epsilon,q)$, i.e.,
\begin{eqnarray}
\bar{N}(\bar{q})&=&N(q) + \tilde{N}(\tilde{q}^2,\epsilon,q).
\end{eqnarray}
Here the $\tilde{N}(\tilde{q}^2,\epsilon,q)$ part is related to $R$:
\begin{eqnarray}
R&\equiv&\int d^d \bar{q}
\frac{\tilde{N}(\tilde{q}^2,\epsilon,q)}{\bar{D}_1 \bar{D}_2
...\bar{D}_N}.
\end{eqnarray}
Feynman rules for the $R$ effective vertices can be obtained from
all possible one-particle irreducible Green functions, which are
enough up to 4 external legs in the SM, accounting for
the ultraviolet nature of the rational terms.

The $d$-dimensional momentum $\bar{q}$, the $d$-dimensional metric
tensor $\bar{g}_{\bar{\mu}\bar{\nu}}$, and the $d$-dimensional Dirac
matrices $\bar{\gamma}_{\bar{\mu}}$ are all split into a
4-dimensional part and a $-2\epsilon$-dimensional part
\begin{eqnarray}
\bar{q}_{\bar{\mu}}=q_{\mu}+\tilde{q}_{\tilde{\mu}},\hspace{0.2cm}
\bar{g}_{\bar{\mu}\bar{\nu}}=g_{\mu\nu}+\tilde{g}_{\tilde{\mu}\tilde{\nu}},\hspace{0.2cm}
\bar{\gamma}_{\bar{\mu}}=\gamma_{\mu}+\tilde{\gamma}_{\tilde{\mu}}.
\end{eqnarray}

Instead of performing supersymmetry-preserving but technically
complicated dimensional reduction (DRED) \cite{Siegel:1979wq}, we
use DREG for convenience. Therefore, the $d$-dimensional metric
tensor $\bar{g}_{\mu\nu}$, the $-2\epsilon$-dimensional tensor
$\tilde{g}_{\mu\nu}$, and the 4-dimensional tensor $g_{\mu\nu}$
satisfy following relations~:
\begin{center}
\begin{eqnarray}
\bar{g}_{\mu\rho}g^{\rho}_{\nu}=g_{\mu\nu},
~\bar{g}_{\mu\rho}\tilde{g}^{\rho}_{\nu}=\tilde{g}_{\mu\nu},
~\tilde{g}_{\mu\rho}g^{\rho}_{\nu}=0,~\bar{g}_{\mu\rho}\bar{g}^{\rho}_{\nu}=\bar{g}_{\mu\nu},\nonumber\\
g_{\mu\rho}g^{\rho}_{\nu}=g_{\mu\nu},~\tilde{g}_{\mu\rho}\tilde{g}^{\rho}_{\nu}=\tilde{g}_{\mu\nu},
~\bar{g}^{\mu}_{\mu}=d,~g^{\mu}_{\mu}=4,~\tilde{g}_{\mu}^{\mu}=-2\epsilon.
\end{eqnarray}
\end{center}
In our regularization we set $d > 4$, so that
\begin{eqnarray}
q^2\rightarrow q^2+\tilde{q}^2.
\end{eqnarray}
To maintain the advantages of the helicity method for loop
calculations, we choose the four dimensional helicity (FDH)
\cite{Bern:1991aq} and 't Hooft-Veltman (HV) schemes
\cite{'tHooft:1972fi} in the present paper
\begin{eqnarray}
R\Bigl |_{HV}&=& \int d^d \bar{q}
\frac{\tilde{N}(\tilde{q}^2,\epsilon,q)}{\bar{D}_1 \bar{D}_2
...\bar{D}_N},\nonumber\\
 R\Bigl |_{FDH}&=& \int {d^d \bar{q}
\frac{\tilde{N}(\tilde{q}^2,\epsilon=0,q)}{\bar{D}_1 \bar{D}_2
...\bar{D}_N}}.
\end{eqnarray}

The last comment in this section refers to our $\gamma_5$
strategies. We choose two schemes here. One scheme is identical to
that used in Refs.\cite{Draggiotis:2009yb,Garzelli:2009is}, which
was proposed by Kreimer, K$\ddot{o}$rner, and Schilcher
\cite{Kreimer:1993bh} and we call it the KKS scheme for brevity, and
the other one is the 't Hoof-Veltman scheme
\cite{'tHooft:1972fi,Bollini:1972bi,Breitenlohner:1977hr}, which
gives new results in the present paper. These two schemes are
algebraically consistent,  at least at one-loop level. In the KKS
scheme, we define a unique generator which anti-commutates with all
other generators in Clifford algebra,
 as  $\gamma_5$. So the
anti-commutation relation $\{\gamma_5,\bar{\gamma}_{\bar{\mu}}\}=0$
is maintained. To obtain the anomaly, we give up the cyclic relation
in the ordinary Dirac trace, and  use a projection relation onto
four dimensional subspace
\begin{center}
\begin{eqnarray}
Tr(\gamma_5 \bar{\gamma}_{\bar{\mu}_{1}} ...
\bar{\gamma}_{\bar{\mu}_{2k}})&\equiv& tr({\cal
P}(\gamma_{5})\bar{\gamma}_{\bar{\mu}_{1}} ...
\bar{\gamma}_{\bar{\mu}_{2k}}),
\end{eqnarray}
\end{center}
where ${\cal P}(\gamma_{5})\equiv\frac{i}{4!}
\hspace{0.2cm}\varepsilon_{\mu_1\mu_2\mu_3\mu_4}
\gamma^{\mu_1}\gamma^{\mu_2}\gamma^{\mu_3}\gamma^{\mu_4}$ with
Lorentz indexes of the totally antisymmetric tensor
$\varepsilon_{\mu_1\mu_2\mu_3\mu_4}$ all in 4 dimensions. To perform
calculations for some specific processes unambiguously, we choose a
unique "special vertex", called the "reading point", in all Feynman
diagrams. All $\gamma_5$'s are anti-commuted to reach the "reading
point" before performing the $d$-dimensional algebra calculation.
This treatment generally produces a term that is proportional to the
total antisymmetric $\varepsilon$ tensor with different "reading
points". The HV scheme defines
$\gamma_{5}\equiv\frac{i}{4!}\hspace{0.2cm}\varepsilon_{\mu_1\mu_2\mu_3\mu_4}
\gamma^{\mu_1}\gamma^{\mu_2}\gamma^{\mu_3}\gamma^{\mu_4}$ in $d$
dimensions. Therefore, the anti-commutation relation is violated by
\begin{center}
\begin{eqnarray}
\{\gamma_5,\gamma_{\mu}\}=0,~[\gamma_5,\tilde{\gamma}_{\tilde{\mu}}]=0.
\end{eqnarray}
\end{center}
This definition explicitly forbids us to deal with covariant
$\gamma$ algebra and often makes calculations very inconvenient.
Nevertheless,  it is to date the only known scheme within DREG that
has been demonstrated to be consistent in all orders. Actually, the
violation of anti-commutation also needs to include an extra finite
renormalization by hand in calculations to restore the Ward
identities. For instance in the $W$ boson decays to hadronic
partons, the Ward identity is spoiled in the HV $\gamma_5$ scheme at
one loop level. The $\mathcal{O}(\alpha_s)$ corrections should
include an axial-vector current renormalization
\begin{center}
\begin{eqnarray}
\Gamma^{ren}_{\mu5}&=&\left(1-\frac{\alpha_s~
C_F}{2\pi}(1+\lambda_{HV})\right)\Gamma^{bare}_{\mu5},
\end{eqnarray}
\end{center}
where $\lambda_{HV}$ is defined in the next section and
$C_F=\frac{N_c^2-1}{2N_c}$ is implicitly understood. This
non-anticommuting $\gamma_5$ scheme was also discussed from the
action principles in Ref.\cite{Martin:1999cc}. Moreover, some issues
about dimensional renormalization with $\gamma_5$ can also be found
in Refs.\cite{Martin:1999cc,Schubert:1988ke}.
\section{Notations and Feynman rules \label{sec:3}}

In this section, we will briefly describe our notations and tree
level Feynman rules that will be used in the next section.

Two parameters, $\lambda_{HV}$ and $g5s$, are introduced in our
formulae to denote the different DREG and $\gamma_5$ schemes. Here,
$\lambda_{HV}=1(0)$ corresponds to the HV (FDH) regularization
scheme, while $g5s=1(-1)$ corresponds to the KKS (HV) $\gamma_5$
scheme. Our notations are as follows: $L_1=e, L_2=\mu, L_3=\tau,
L_4=\nu_e, L_5=\nu_{\mu}, L_6=\nu_{\tau}, Q_1=u, Q_2=d, Q_3=s,
Q_4=c, Q_5=b, Q_6=t$. In addition,
$e_1=e,e_2=\mu,e_3=\tau,\nu_1=\nu_e,\nu_2=\nu_{\mu},\nu_3=\nu_{\tau},U_1=u,U_2=c,U_3=t,D_1=d,D_2=s,D_3=b$.
The vector boson fields are denoted by $A$,$Z$ and $W^{\pm}$ with
the generic symbol $V$, and  the physical scalar Higgs field and the
scalar goldstone bosons are written as $H$, $\phi^0$, and
$\phi^{\pm}$, respectively. Fermions are also generically symbolized
by $F$, while the mass, charge, and the third isospin component of
$F$ are denoted by $m_F$, $Q_F$, and $I_{3F}$, respectively.   $c_w$
and $s_w$ respectively denote  sine and cosine of the Weinberg
angle. $N_c$ is the number of colors and $V_{u_i,d_j}$ refers to the
Cabibbo-Kobayashi-Maskawa (CKM)
 matrix elements.
$\Omega^{\pm}\equiv\frac{1\pm\gamma_5}{2}$ are the chiral projector
operators. Finally, $g_s$ is the coupling constant of strong
interaction, while $e$ is the coupling constant of QED.

As we use FeynArts \cite{Kublbeck:1990xc} to generate all Feynman
amplitudes, our conventions for tree level Feynman rules follow
Ref.\cite{Denner:1991kt}. Due to the violation of anti-commutation
relation in HV $\gamma_5$ scheme, i.e.
$\{\gamma_5,\bar{\gamma}_{\bar{\mu}}\}\neq0$, some modifications to
conventional Feynman rules are introduced. For coupling to
left-handed fields the symmetrically defined vertex
~$\frac{1}{2}(1+\gamma_5)\bar{\gamma}_{\bar{\mu}}(1-\gamma_5)=\gamma_{\mu}(1-\gamma_5)$,
instead of $\bar{\gamma}_{\bar{\mu}}(1-\gamma_5)$, should be
used~\cite{Korner:1989is}. Especially, the tree level Feynman rules
of vertices $F\bar{F}Z$ should be modified as follows
\begin{center}
\begin{eqnarray}
\frac{ie}{c_ws_w}\left(I_{3F}\bar{\gamma}_{\bar{\mu}}\Omega^--s_w^2Q_F\bar{\gamma}_{\bar{\mu}}\right)
\hspace{0.2cm}\rightarrow\hspace{0.2cm}
\frac{ie}{c_ws_w}\left(I_{3F}\gamma_{\mu}\Omega^--s_w^2Q_F\bar{\gamma}_{\bar{\mu}}\right),
\end{eqnarray}
\end{center}
and similarly to $F_1\bar{F}_2W^{\pm}$ vertices.
\section{Results \label{sec:4}}

In this section, our results in 't Hooft-Feynman gauge are
presented. It is easy to check that if one sets $g5s=1$, the results
in \cite{Draggiotis:2009yb,Garzelli:2009is} are all recovered.
\subsection{Effective vertices in QCD}

In this subsection, we give a complete list of all non-vanishing $R$
effective vertices in QCD, where all the internal lines in
one-particle irreducible diagrams represent QCD particles.
\subsubsection{QCD effective vertices with 2 external legs}

\begin{center}
\begin{figure}
\hspace{0cm}\includegraphics[width=9cm]{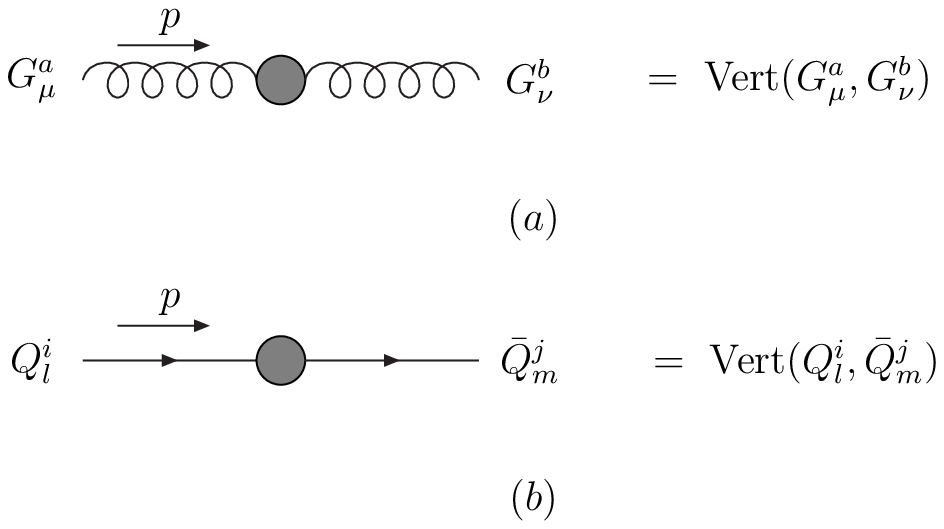}
\caption{\label{fig:qcd2} All possible non-vanishing 2-point
vertices in QCD.}
\end{figure}
\end{center}

All possible non-vanishing 2-point vertices in QCD are shown in
Fig.\ref{fig:qcd2}.

\begin{center}
\textbf{{\small Gluon-Gluon vertex}}
 \vspace{0.1cm}
\end{center}

The corresponding effective vertex is shown in
Fig.\ref{fig:qcd2}~$(a)$ with the following expression
\begin{center}
\begin{eqnarray}
{\rm Vert}(G^{a}_{\mu},G^{b}_{\nu})&=& \frac{i g_s^2N_c}{48
\pi^2}~\delta^{ab}~\left[\frac{p^2}{2}g_{\mu\nu}+\lambda_{HV}\left(g_{\mu\nu}p^2-p_{\mu}p_{\nu}\right)
+\sum_{Q}\frac{p^2-6m_Q^2}{N_c}g_{\mu\nu}\right].
\end{eqnarray}
\end{center}
\vspace{0.5cm}

\begin{center}
\textbf{{\small Quark-Quark vertices}}
 \vspace{0.4cm}
\end{center}

The corresponding diagram is shown in Fig.\ref{fig:qcd2}~$(b)$ with
the following expression
\begin{center}
\begin{eqnarray}
{\rm Vert}(Q^i_l,\bar{Q}_m^j)&=& \frac{i g_s^2}{16
\pi^2}~\frac{N_c^2-1}{2N_c}~\delta^{ij}~\delta_{lm}~\left(-\rlap/p+2m_{Q_l}\right)\lambda_{HV}.
\end{eqnarray}
\end{center}
\vspace{0.5cm}

\subsubsection{QCD effective vertices with 3 external legs}

\begin{center}
\begin{figure}
\hspace{0cm}\includegraphics[width=7cm]{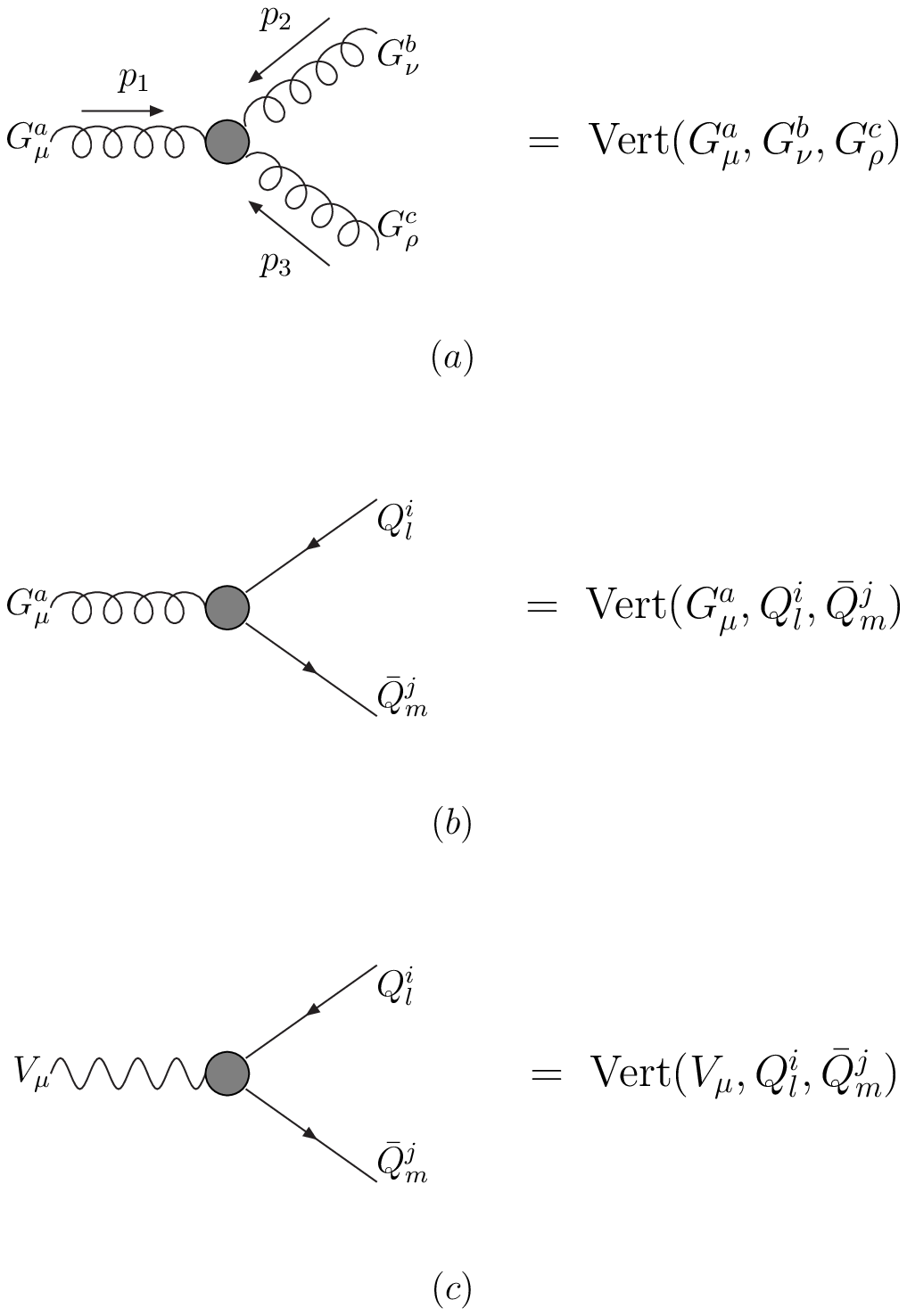}
\hspace{0.2cm}\includegraphics[width=7cm]{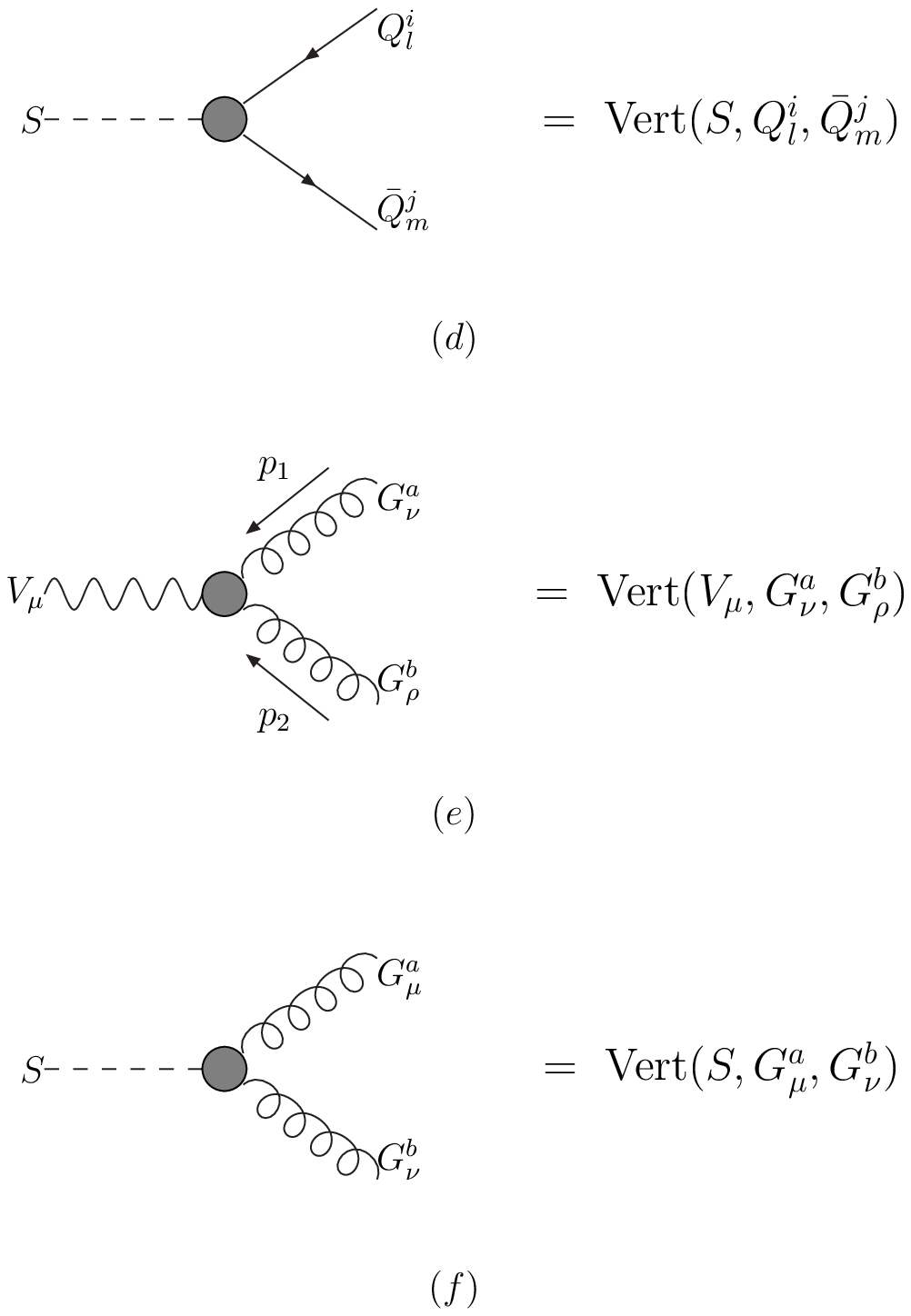}
\caption{\label{fig:qcd3} All possible non-vanishing 3-point
vertices in QCD.}
\end{figure}
\end{center}

All possible non-vanishing 3-point vertices in QCD are shown in
Fig.\ref{fig:qcd3}.

\begin{center}
\textbf{{\small Gluon-Gluon-Gluon vertex}}
 \vspace{0.1cm}
\end{center}

The corresponding diagram is shown in Fig.\ref{fig:qcd3}~$(a)$ with
the following expression for $N_f$-flavor quarks
\begin{center}
\begin{eqnarray}
{\rm Vert}(G^a_{\mu},G^b_{\nu},G^c_{\rho})&=& -\frac{g_s^3N_c}{48
\pi^2}~\left(\frac{7}{4}+\lambda_{HV}+\frac{2N_f}{N_c}\right)f^{abc}~V_{\mu\nu\rho}(p_1,p_2,p_3),
\end{eqnarray}
\end{center}
where
\begin{center}
\begin{eqnarray}
V_{\mu\nu\rho}(p_1,p_2,p_3)&=&~g_{\mu\nu}(p_2-p_1)_{\rho}+g_{\nu\rho}(p_3-p_2)_{\mu}+g_{\rho\mu}(p_1-p_3)_{\nu}.
\end{eqnarray}
\end{center}
\vspace{0.5cm}

\begin{center}
\textbf{{\small Gluon-Quark-Quark vertex}}
 \vspace{0.1cm}
\end{center}

The corresponding diagram is shown in Fig.\ref{fig:qcd3}~$(b)$ with
the following expression
\begin{center}
\begin{eqnarray}
{\rm Vert}(G^a_{\mu},Q^i_l,\bar{Q}^j_m)&=&\delta_{lm}~\frac{i
g_s^3}{16
\pi^2}~T^a_{ji}~\frac{N_c^2-1}{2N_c}~\gamma_{\mu}\left(1+\lambda_{HV}\right).
\end{eqnarray}
\end{center}
\vspace{0.5cm}

\begin{center}
\textbf{{\small Vector-Quark-Quark vertices}}
 \vspace{0.1cm}
\end{center}

A typical Vector-Quark-Quark vertex is shown in
Fig.\ref{fig:qcd3}~$(c)$ with the following expression
\begin{center}
\begin{eqnarray}
{\rm Vert}(V_{\mu},Q^i_l,\bar{Q}^j_m)&=&-\frac{g_s^2}{16
\pi^2}~\frac{N_c^2-1}{2N_c}~\delta^{ij}~\left(1+\lambda_{HV}\right)\nonumber\\&&
~\gamma_{\mu}\left(v^{VQ_l\bar{Q}_m}
+g5s~a^{VQ_l\bar{Q}_m}\gamma_5\right).\label{eq:wqq}
\end{eqnarray}
\end{center}
The actual values of $V$, $Q_l$, $\bar{Q}_m$, $v^{VQ_l\bar{Q}_m}$
and $a^{VQ_l\bar{Q}_m}$ are
\begin{center}
\begin{eqnarray}
v^{AQ_l\bar{Q}_m}&=&-ieQ_{Q_l}\delta_{lm},\nonumber\\
a^{AQ_l\bar{Q}_m}&=&0,\nonumber\\
v^{ZQ_l\bar{Q}_m}&=&\frac{ies_w}{c_w}\left(\frac{I_{3Q_l}}{2s_w^2}-Q_{Q_l}\right)\delta_{lm},\nonumber\\
a^{ZQ_l\bar{Q}_m}&=&-\frac{ieI_{3Q_l}}{2s_wc_w}\delta_{lm},\nonumber\\
v^{W^-U_l\bar{D}_m}&=&V_{D_m,U_l}^{\dagger}\frac{ie}{2\sqrt{2}s_w},\nonumber\\
a^{W^-U_l\bar{D}_m}&=&-V_{D_m,U_l}^{\dagger}\frac{ie}{2\sqrt{2}s_w},\nonumber\\
v^{W^+D_l\bar{U}_m}&=&V_{U_m,D_l}\frac{ie}{2\sqrt{2}s_w},\nonumber\\
a^{W^+D_l\bar{U}_m}&=&-V_{U_m,D_l}\frac{ie}{2\sqrt{2}s_w}.\label{eq:vqq}
\end{eqnarray}
\end{center}
\vspace{0.5cm}

\begin{center}
\textbf{{\small Scalar-Quark-Quark vertices}}
 \vspace{0.1cm}
\end{center}

The generic diagram is shown in Fig.\ref{fig:qcd3}~$(d)$ with the
following expression
\begin{center}
\begin{eqnarray}
{\rm Vert}(S,Q^i_l,\bar{Q}^j_m)&=&-\frac{g_s^2}{8
\pi^2}~\frac{N_c^2-1}{2N_c}~\delta^{ij}~\left(1+\lambda_{HV}\right)\nonumber\\&&
~\left(v^{SQ_l\bar{Q}_m}
+g5s~a^{SQ_l\bar{Q}_m}\gamma_5\right).
\end{eqnarray}
\end{center}
The actual values of $S$, $Q_l$, $\bar{Q}_m$, $v^{SQ_l\bar{Q}_m}$
and $a^{SQ_l\bar{Q}_m}$ are
\begin{center}
\begin{eqnarray}
v^{HQ_l\bar{Q}_m}&=&-\delta_{lm}~\frac{i~e~m_{Q_l}}{2m_Ws_w},
~~~~~~~~~a^{HQ_l\bar{Q}_m}=0,\nonumber\\
v^{\phi^0U_l\bar{U}_m}&=&0,
~~~~~~~~~~~~~~~~~~~~~~~~~~a^{\phi^0U_l\bar{U}_m}=-\delta_{lm}~\frac{e~m_{U_l}}{2m_Ws_w},\nonumber\\
v^{\phi^0D_l\bar{D}_m}&=&0,
~~~~~~~~~~~~~~~~~~~~~~~~~~a^{\phi^0D_l\bar{D}_m}=\delta_{lm}~\frac{e~m_{D_l}}{2m_Ws_w},\nonumber\\
v^{\phi^-U_l\bar{D}_m}&=&V^{\dagger}_{D_m,U_l}\frac{i~e}{2\sqrt{2}m_Ws_w}\left(m_{U_l}-m_{D_m}\right),\nonumber\\
a^{\phi^-U_l\bar{D}_m}&=&V^{\dagger}_{D_m,U_l}\frac{i~e}{2\sqrt{2}m_Ws_w}\left(m_{U_l}+m_{D_m}\right),\nonumber\\
v^{\phi^+D_l\bar{U}_m}&=&V_{U_m,D_l}\frac{i~e}{2\sqrt{2}m_Ws_w}\left(m_{U_m}-m_{D_l}\right),\nonumber\\
a^{\phi^+D_l\bar{U}_m}&=&-V_{U_m,D_l}\frac{i~e}{2\sqrt{2}m_Ws_w}\left(m_{U_m}+m_{D_l}\right).
\label{eq:sqq}
\end{eqnarray}
\end{center}
\vspace{0.5cm}

\begin{center}
\textbf{{\small Vector-Gluon-Gluon vertices}}
 \vspace{0.1cm}
\end{center}

The generic diagram is shown in Fig.\ref{fig:qcd3}~$(e)$ with the
following expression
\begin{center}
\begin{eqnarray}
{\rm Vert}(V_{\mu},G^a_{\nu},G^b_{\rho})&=& -\frac{ig_s^2}{12
\pi^2}~\delta^{ab}~\varepsilon_{\mu\nu\rho\left(p_1-p_2\right)}~\sum_{l}a^{VQ_l\bar{Q}_l},
\end{eqnarray}
\end{center}
with all of the expressions of $a^{VQ_l\bar{Q}_l}$ given in
Eq.(\ref{eq:vqq}). \vspace{0.5cm}

\begin{center}
\textbf{{\small Scalar-Gluon-Gluon vertices}}
 \vspace{0.1cm}
\end{center}

The generic diagram is shown in Fig.\ref{fig:qcd3}~$(f)$ with the
following expression
\begin{center}
\begin{eqnarray}
{\rm Vert}(S,G^a_{\mu},G^b_{\nu})&=& \frac{g_s^2}{8
\pi^2}~\delta^{ab}~g_{\mu\nu}~\sum_{l}\left(v^{SQ_l\bar{Q}_l}m_{Q_l}\right),
\end{eqnarray}
\end{center}
with all of the expressions of $v^{SQ_l\bar{Q}_l}$ given in
Eq.(\ref{eq:sqq}). \vspace{0.5cm}

\subsubsection{QCD effective vertices with 4 external legs}

\begin{center}
\begin{figure}
\hspace{0cm}\includegraphics[width=7cm]{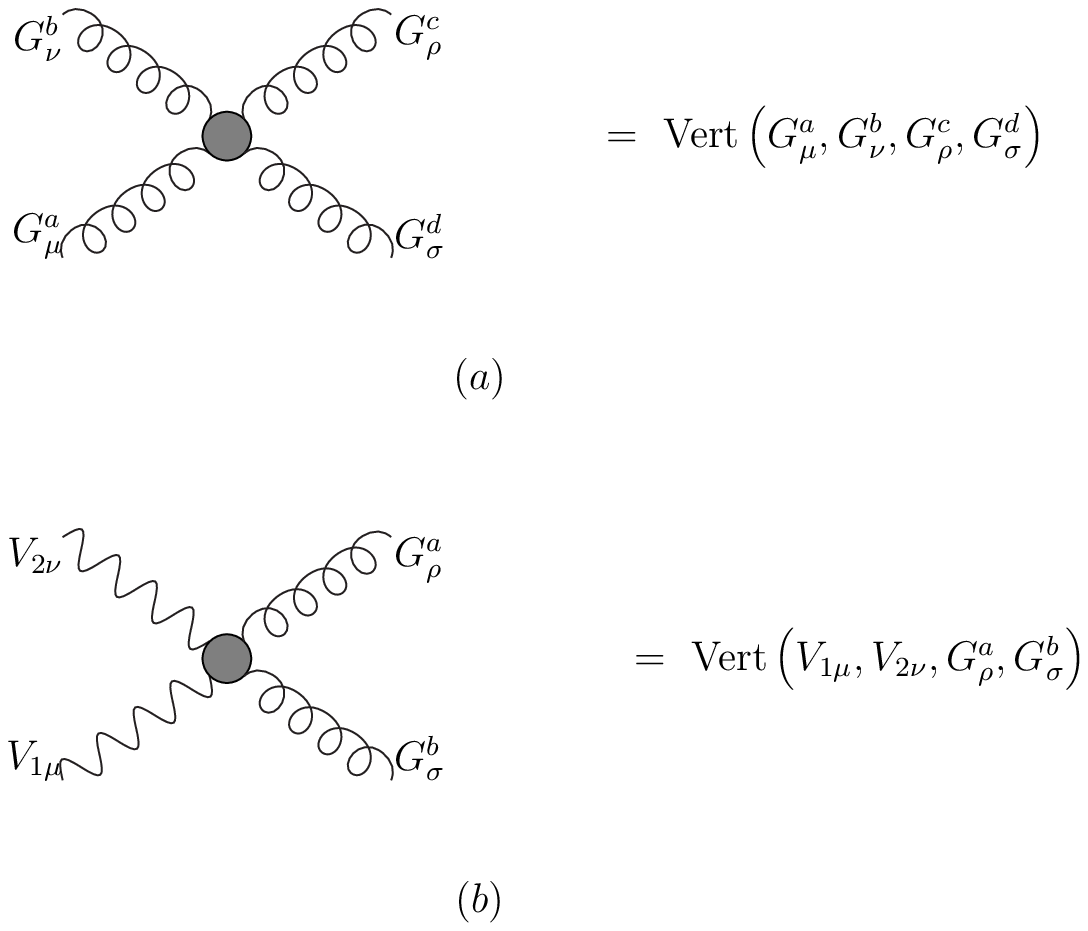}
\hspace{0cm}\includegraphics[width=7cm]{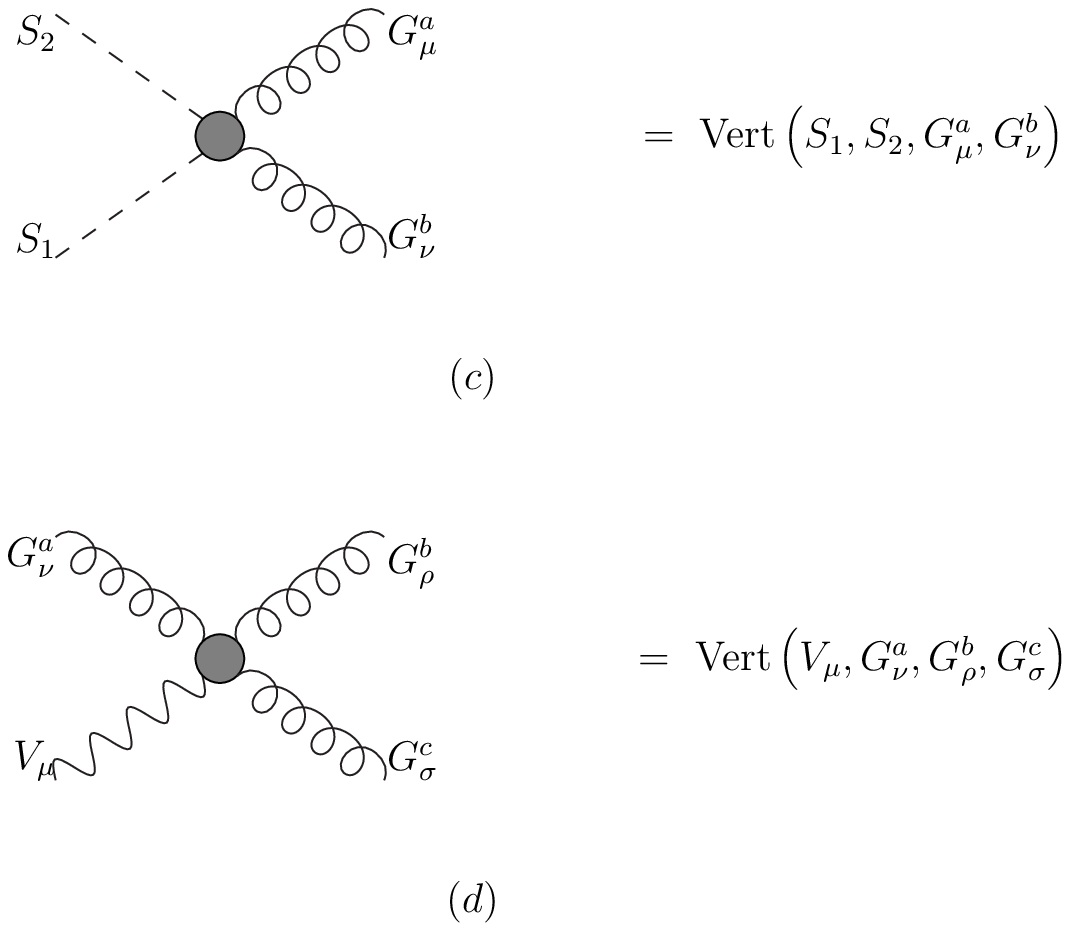}
\caption{\label{fig:qcd4} All possible non-vanishing 4-point
vertices in QCD.}
\end{figure}
\end{center}

All possible non-vanishing 4-point vertices in QCD are shown in
Fig.\ref{fig:qcd4}.

\begin{center}
\textbf{{\small Gluon-Gluon-Gluon-Gluon vertex}}
 \vspace{0.1cm}
\end{center}

The corresponding diagram is shown in Fig.\ref{fig:qcd4}~$(a)$ with
the following expression for $N_f$-flavor quarks
\begin{center}
\begin{eqnarray}
{\rm
Vert}\left(G^a_{\mu},G^b_{\nu},G^c_{\rho},G^d_{\sigma}\right)&=&
\frac{i g_s^4}{48
\pi^2}~\left(C_1~g_{\mu\nu}g_{\rho\sigma}+C_2~g_{\mu\rho}g_{\nu\sigma}+C_3~g_{\mu\sigma}g_{\nu\rho}\right),
\end{eqnarray}
\end{center}
where
\begin{center}
\begin{eqnarray}
C_1&=&Tr(\{T^a,T^b\}\{T^c,T^d\})~\left(5N_c+2\lambda_{HV}N_c+6N_f\right)\nonumber\\&&
-\left(Tr(T^aT^cT^bT^d)+Tr(T^aT^dT^bT^c)\right)\left(12N_c+4\lambda_{HV}N_c+10N_f\right)~\nonumber\\&&
-\left(\delta^{ab}\delta^{cd}+\delta^{ac}\delta^{bd}+\delta^{ad}\delta^{bc}\right),\nonumber\\
C_2&=&C_1(b\leftrightarrow
c)=Tr(\{T^a,T^c\}\{T^b,T^d\})~\left(5N_c+2\lambda_{HV}N_c+6N_f\right)\nonumber\\&&
-\left(Tr(T^aT^bT^cT^d)+Tr(T^aT^dT^cT^b)\right)\left(12N_c+4\lambda_{HV}N_c+10N_f\right)~\nonumber\\&&
-\left(\delta^{ab}\delta^{cd}+\delta^{ac}\delta^{bd}+\delta^{ad}\delta^{bc}\right),\nonumber\\
C_3&=&C_1(b\leftrightarrow
d)=Tr(\{T^a,T^d\}\{T^c,T^b\})~\left(5N_c+2\lambda_{HV}N_c+6N_f\right)\nonumber\\&&
-\left(Tr(T^aT^cT^dT^b)+Tr(T^aT^bT^dT^c)\right)\left(12N_c+4\lambda_{HV}N_c+10N_f\right)~\nonumber\\&&
-\left(\delta^{ab}\delta^{cd}+\delta^{ac}\delta^{bd}+\delta^{ad}\delta^{bc}\right).
\end{eqnarray}
\end{center}
\vspace{0.5cm}

\begin{center}
\textbf{{\small Vector-Vector-Gluon-Gluon vertices}}
 \vspace{0.1cm}
\end{center}

The generic diagram is shown in Fig.\ref{fig:qcd4}~$(b)$ with the
following expression
\begin{center}
\begin{eqnarray}
{\rm Vert}\left(V_{1\mu},V_{2\nu},G^a_{\rho},G^b_{\sigma}\right)&=&
-\frac{i
g_s^2}{24\pi^2}~\delta^{ab}~\sum_{l,m}\left[\left(v^{V_1Q_l\bar{Q}_m}v^{V_2Q_m\bar{Q}_l}
+a^{V_1Q_l\bar{Q}_m}a^{V_2Q_m\bar{Q}_l}\right)\right.\nonumber\\&&
\left(g_{\mu\nu}g_{\rho\sigma}+g_{\mu\rho}g_{\nu\sigma}+g_{\mu\sigma}g_{\nu\rho}\right)+\frac{1-g5s}{2}
~8a^{V_1Q_l\bar{Q}_m}a^{V_2Q_m\bar{Q}_l}\nonumber\\&&\left.g_{\mu\nu}g_{\rho\sigma}\right].
\end{eqnarray}
\end{center}
All of the expressions for $v^{VQ_l\bar{Q}_m}$,$a^{VQ_l\bar{Q}_m}$
are given in Eq.(\ref{eq:vqq}). \vspace{0.5cm}

\begin{center}
\textbf{{\small Scalar-Scalar-Gluon-Gluon vertices}}
 \vspace{0.1cm}
\end{center}

The generic diagram is shown in Fig.\ref{fig:qcd4}~$(c)$ with the
following expression
\begin{center}
\begin{eqnarray}
{\rm Vert}\left(S_1,S_2,G^a_{\mu},G^b_{\nu}\right)&=& \frac{i
g_s^2}{24
\pi^2}~\delta^{ab}~\sum_{l,m}\left[3\left(v^{S_1Q_l\bar{Q}_m}v^{S_2Q_m\bar{Q}_l}-
a^{S_1Q_l\bar{Q}_m}a^{S_2Q_m\bar{Q}_l}\right)\right.\nonumber\\&&\left.
-\frac{1-g5s}{2}~8a^{S_1Q_l\bar{Q}_m}a^{S_2Q_m\bar{Q}_l}\right]~g_{\mu\nu}.
\end{eqnarray}
\end{center}
All of the expressions for $v^{SQ_l\bar{Q}_m}$,$a^{SQ_l\bar{Q}_m}$
are given in Eq.(\ref{eq:sqq}).
 \vspace{0.5cm}

\begin{center}
\textbf{{\small Vector-Gluon-Gluon-Gluon vertices}}
 \vspace{0.1cm}
\end{center}

The generic diagram is shown in Fig.\ref{fig:qcd4}~$(d)$ with the
following expression
\begin{center}
\begin{eqnarray}
{\rm Vert}\left(V_{\mu},G^a_{\nu},G^b_{\rho},G^c_{\sigma}\right)&=&
-\frac{
g_s^3}{24\pi^2}~\sum_{l}\left[Tr(T^a\{T^b,T^c\})~v^{VQ_l\bar{Q}_l}\left(g_{\mu\nu}g_{\rho\sigma}
+g_{\mu\rho}g_{\nu\sigma}+g_{\mu\sigma}g_{\nu\rho}\right)\right.\nonumber\\&&\left.
-i9~Tr(T^a[T^b,T^c])~a^{VQ_l\bar{Q}_l}\varepsilon_{\mu\nu\rho\sigma}\right],
\end{eqnarray}
\end{center}
with $\{A,B\}\equiv AB+BA$ and $[A,B]\equiv AB-BA$.~All of the
expressions for $v^{VQ_l\bar{Q}_l}$,$a^{VQ_l\bar{Q}_l}$ are given in
Eq.(\ref{eq:vqq}). \vspace{0.5cm}

\subsection{Effective vertices in Electroweak part}

A complete list of the non-vanishing $R$ effective vertices in the
Electroweak part is given below.

\subsubsection{Electroweak effective vertices with 2 external legs}

\begin{center}
\begin{figure}
\hspace{0cm}\includegraphics[width=9cm]{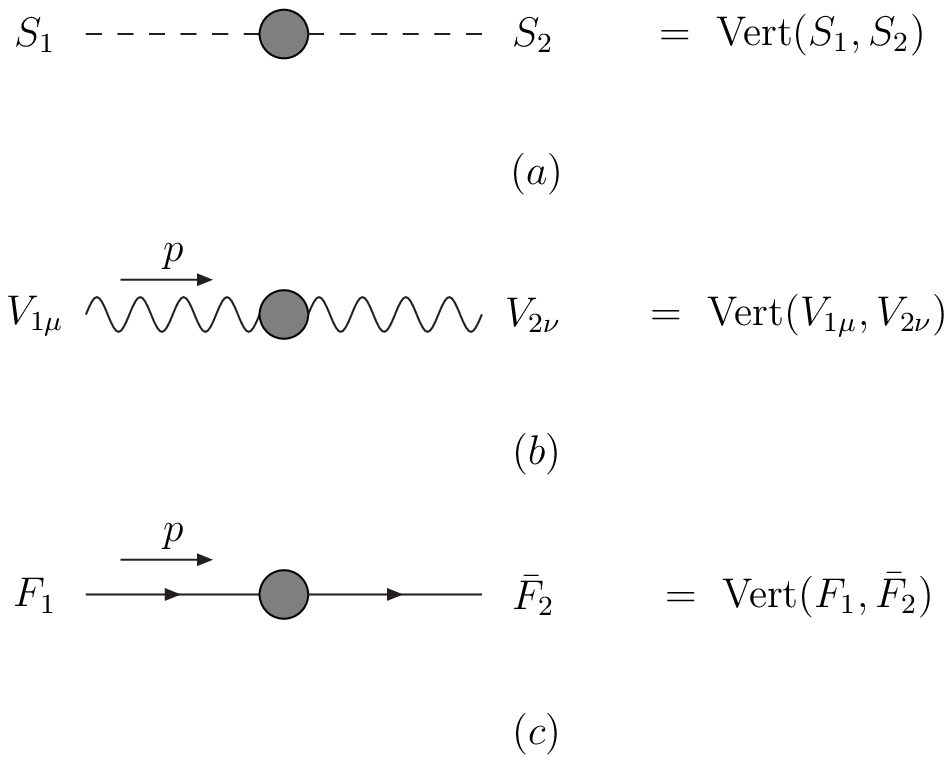}
\caption{\label{fig:ew2} All possible non-vanishing 2-point vertices
in EW.}
\end{figure}
\end{center}

All possible non-vanishing 2-point vertices in Electroweak (EW) are
shown in Fig.\ref{fig:ew2}.

\begin{center}
\textbf{{\small Scalar-Scalar vertices}}
 \vspace{0.1cm}
\end{center}

The generic effective vertex is shown in Fig.\ref{fig:ew2}~$(a)$
with the following expression
\begin{center}
\begin{eqnarray}
{\rm Vert}\left(S_1,S_2\right)&=& \frac{i e^2}{16
\pi^2s_w^2}~C^{S_1S_2},
\end{eqnarray}
\end{center}
with the actual value of $S_1$, $S_2$ and $C^{S_1S_2}$
\begin{center}
\begin{eqnarray}
C^{H\phi^0}&=&0,\nonumber\\
C^{HH}&=&\frac{1-6\lambda_{HV}}{2}\left(1+\frac{1}{2c_w^4}\right)m_W^2-\left(1+\frac{1}{2c_w^2}\right)\frac{p^2}{12}+K,
\nonumber\\
C^{\phi^0\phi^0}&=&\frac{m_W^2}{4}+\frac{m_H^2}{8c_w^2}+\frac{1-4\lambda_{HV}}{4}\left(1+\frac{1}{2c_w^4}\right)m_W^2\nonumber\\&&
-\left(1+\frac{1}{2c_w^2}\right)\frac{p^2}{12}+g5s~K,\nonumber\\
C^{\phi^-\phi^+}&=&\frac{m_W^2}{4c_w^2}+\frac{m_H^2}{8}+\frac{\left(3-4\lambda_{HV}\right)c_w^4-2c_w^2
+\left(\frac{1}{2}-2\lambda_{HV}\right)}{4c_w^4}m_W^2\nonumber\\&&
-\left(1+\frac{1}{2c_w^2}\right)\frac{p^2}{12}+\frac{1}{2m_W^2}\left\{\frac{1+g5s}{2}~\sum_{l}\left(m_{e_l}^2
\left(m_{e_l}^2-\frac{p^2}{3}\right)\right)\right.\nonumber\\&&
+N_c\sum_{l,m}\left[V_{U_l,D_m}V^{\dagger}_{D_m,U_l}\left(\frac{1+g5s}{2}~\left(m_{U_l}^2+m_{D_m}^2\right)\right.\right.
\nonumber\\&&\left.\left.\left.
-\frac{1-g5s}{2}~2m_{U_l}m_{D_m}\right)\left(m_{U_l}^2+m_{D_m}^2-\frac{p^2}{3}\right)\right]\right\},
\end{eqnarray}
\end{center}
where
\begin{center}
\begin{eqnarray}
K&=&\sum_{l}\left[\frac{m_{e_l}^2}{m_W^2}\left(m_{e_l}^2-\frac{p^2}{6}\right)\right]
+N_c\sum_{l}\left[\frac{m_{Q_l}^2}{m_W^2}\left(m_{Q_l}^2-\frac{p^2}{6}\right)\right].
\end{eqnarray}
\end{center}
To compare our expressions with that in Ref.\cite{Garzelli:2009is},
we use $m_{\phi^0}^2=m_{Z}^2=\frac{m_W^2}{c_w^2}$ and
$m_{\phi^{\pm}}^2=m_W^2$ in above expressions and fix the masses of
neutrinos to be zero.
 \vspace{0.5cm}

\begin{center}
\textbf{{\small Vector-Vector vertices}}
 \vspace{0.1cm}
\end{center}

The generic effective vertex is shown in Fig.\ref{fig:ew2}~$(b)$
with the following expression
\begin{center}
\begin{eqnarray}
{\rm Vert}\left(V_{1\mu},V_{2\nu}\right)&=& \frac{i e^2}{
\pi^2}~\left(C_1^{V_1V_2}~p_{\mu}p_{\nu}+C_2^{V_1V_2}~g_{\mu\nu}\right),
\end{eqnarray}
\end{center}
with the  actual values of $V_1$, $V_2$, $C_1^{V_1V_2}$ and
$C_2^{V_1V_2}$
\begin{center}
\begin{eqnarray*}
C_1^{AA}&=&-\frac{\lambda_{HV}}{24},\nonumber\\
C_2^{AA}&=&\frac{1}{8}\left[\frac{\left(1+2\lambda_{HV}\right)p^2}{6}-m_W^2\right]
-\frac{1}{4}\left\{\sum_{l}\left[Q_{L_l}^2\left(m_{L_l}^2-\frac{p^2}{6}\right)\right]
\right.\nonumber\\&&\left.+N_c\sum_{l}\left[Q_{Q_l}^2\left(m_{Q_l}^2-\frac{p^2}{6}\right)\right]\right\},\nonumber\\
C_1^{AZ}&=&\frac{c_w\lambda_{HV}}{24s_w},\nonumber\\
C_2^{AZ}&=&-\frac{c_w}{8s_w}\left[\frac{\left(1+2\lambda_{HV}\right)p^2}{6}-m_W^2\right]+\frac{1}{4c_w}
\left\{\sum_{l}\left[\left(\frac{Q_{L_l}I_{3L_l}}{2s_w}-Q_{L_l}^2s_w\right)\right.\right.\nonumber\\&&
\left.\left.\left(m_{L_l}^2-\frac{p^2}{6}\right)\right]+N_c\sum_{l}\left[\left(\frac{Q_{Q_l}I_{3Q_l}}{2s_w}
-Q_{Q_l}^2s_w\right)\left(m_{Q_l}^2-\frac{p^2}{6}\right)\right]\right\},\nonumber\\
C_1^{ZZ}&=&-\frac{c_w^2\lambda_{HV}}{24s_w^2},\nonumber\\
C_2^{ZZ}&=&\frac{c_w^2}{8s_w^2}\left[\frac{\left(1+2\lambda_{HV}\right)p^2}{6}-m_W^2\right]+\frac{1}{4c_w^2}
\left\{\sum_{l}\left[\left(Q_{L_l}I_{3L_l}-\frac{1+g5s}{2}~\frac{I_{3L_l}^2}{2s_w^2}\right.\right.\right.\nonumber\\&&
\left.\left.-Q_{L_l}^2s_w^2\right)
\left(m_{L_l}^2-\frac{p^2}{6}\right)\right]+N_c\sum_{l}\left[\left(Q_{Q_l}I_{Q_l}-\frac{1+g5s}{2}~
\frac{I_{3Q_l}^2}{2s_w^2}-Q_{Q_l}^2s_w^2 \right)\right.\nonumber\\&&
\left.\left.\left(m_{Q_l}^2-\frac{p^2}{6}\right)\right]\right\},\nonumber\\
C_1^{W^-W^+}&=&-\frac{\lambda_{HV}}{24s_w^2},
\end{eqnarray*}
\end{center}
\begin{center}
\begin{eqnarray}
C_2^{W^-W^+}&=&\frac{1}{8s_w^2}\left[\frac{\left(1+2\lambda_{HV}\right)p^2}{6}-m_W^2\right]-\frac{1+g5s}{2}~
\frac{1}{32s_w^2}\left\{
\sum_{l}\left(m_{e_l}^2-\frac{p^2}{3}\right)\right.\nonumber\\&&\left.+N_c\sum_{l,m}
\left[V_{U_l,D_m}V^{\dagger}_{D_m,U_l}\left(m_{U_l}^2+m_{D_m}^2-\frac{p^2}{3}\right)\right]
\right\}.
\end{eqnarray}
\end{center}
\vspace{0.5cm}

\begin{center}
\textbf{{\small Fermion-Fermion vertices}}
 \vspace{0.1cm}
\end{center}

The generic effective vertex is shown in Fig.\ref{fig:ew2}~$(c)$
with the following expression
\begin{center}
\begin{eqnarray}
{\rm Vert}\left(F_1,\bar{F}_2\right)&=& \frac{i
e^2}{\pi^2}~\left[\left(C^{F_1\bar{F}_2}_{-}\Omega^-+C^{F_1\bar{F}_2}_{+}\Omega^{+}\right)\rlap/p+C_0^{F_1\bar{F}_2}
\right]~\lambda_{HV},
\end{eqnarray}
\end{center}
with the actual values of $F_1$,$\bar{F}_2$, $C^{F_1\bar{F}_2}_{-}$,
$C^{F_1\bar{F}_2}_{+}$ and $C_0^{F_1\bar{F}_2}$
\begin{center}
\begin{eqnarray*}
C^{U_l\bar{U}_m}_{-}&=&\delta_{lm}~\frac{Q_{U_l}^2}{16c_w^2},\nonumber\\
C^{U_l\bar{U}_m}_{+}&=&\delta_{lm}~\frac{1}{16}\left[\frac{1+g5s}{2}~\frac{I_{3U_l}^2}{s_w^2c_w^2}
-\frac{1+g5s}{2}~\frac{2Q_{U_l}I_{3U_l}}{c_w^2}+
\frac{Q_{U_l}^2}{c_w^2}\right.\nonumber\\&&
\left.+\frac{1+g5s}{2}~\frac{1}{2s_w^2}\sum_{g}\left(V_{U_l,D_g}V^{\dagger}_{D_g,U_l}\right)\right],\nonumber\\
C^{U_l\bar{U}_m}_{0}&=&\delta_{lm}~\frac{m_{U_l}Q_{U_l}}{8c_w^2}\left(Q_{U_l}-\frac{1+g5s}{2}~I_{3U_l}\right),\nonumber\\
C^{D_l\bar{D}_m}_{-}&=&\delta_{lm}~\frac{Q_{D_l}^2}{16c_w^2},\nonumber\\
C^{D_l\bar{D}_m}_{+}&=&\delta_{lm}~\frac{1}{16}\left[\frac{1+g5s}{2}~\frac{I_{3D_l}^2}{s_w^2c_w^2}
-\frac{1+g5s}{2}~\frac{2Q_{D_l}I_{3D_l}}{c_w^2}+
\frac{Q_{D_l}^2}{c_w^2}\right.\nonumber\\&&
\left.+\frac{1+g5s}{2}~\frac{1}{2s_w^2}\sum_{g}\left(V_{U_g,D_l}V^{\dagger}_{D_l,U_g}\right)\right],\nonumber\\
C^{D_l\bar{D}_m}_{0}&=&\delta_{lm}~\frac{m_{D_l}Q_{D_l}}{8c_w^2}\left(Q_{D_l}-\frac{1+g5s}{2}~I_{3D_l}\right),
\end{eqnarray*}
\end{center}
\begin{center}
\begin{eqnarray}
C^{L_l\bar{L}_m}_{-}&=&\delta_{lm}~\frac{Q_{L_l}^2}{16c_w^2},\nonumber\\
C^{L_l\bar{L}_m}_{+}&=&\delta_{lm}~\frac{1}{16}\left(\frac{1+g5s}{2}~\frac{I_{3L_l}^2}{s_w^2c_w^2}
-\frac{1+g5s}{2}~\frac{2Q_{L_l}I_{3L_l}}{c_w^2}+
\frac{Q_{L_l}^2}{c_w^2}\right.\nonumber\\&&
\left.+\frac{1+g5s}{2}~\frac{1}{2s_w^2}\right),\nonumber\\
C^{L_l\bar{L}_m}_{0}&=&\delta_{lm}~\frac{m_{L_l}Q_{L_l}}{8c_w^2}\left(Q_{L_l}-\frac{1+g5s}{2}~I_{3L_l}\right).
\end{eqnarray}
\end{center}
\vspace{0.5cm}

\subsubsection{Electroweak effective vertices with 3 external legs}

\begin{center}
\begin{figure}
\hspace{0cm}\includegraphics[width=7cm]{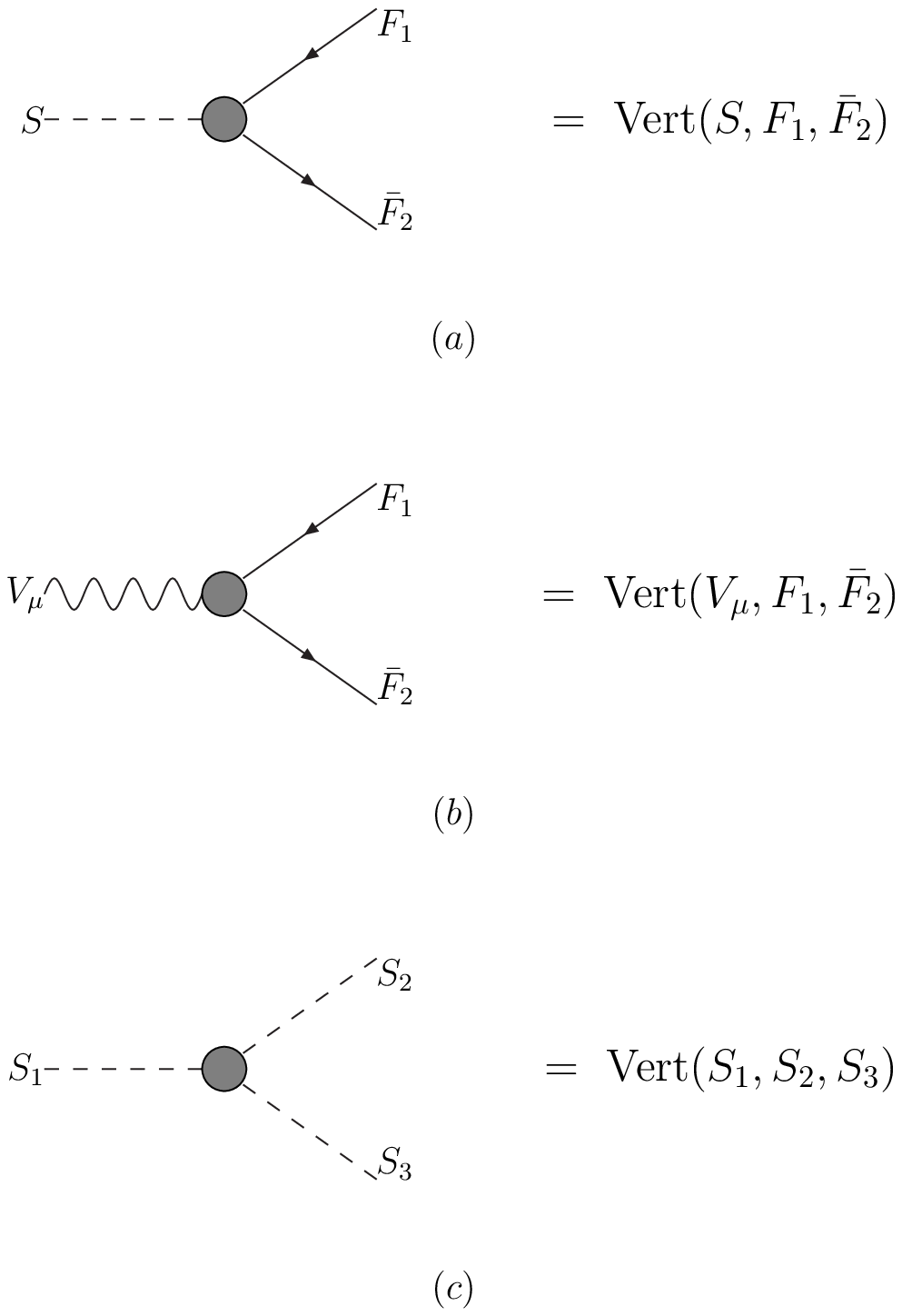}
\hspace{0.2cm}\includegraphics[width=7cm]{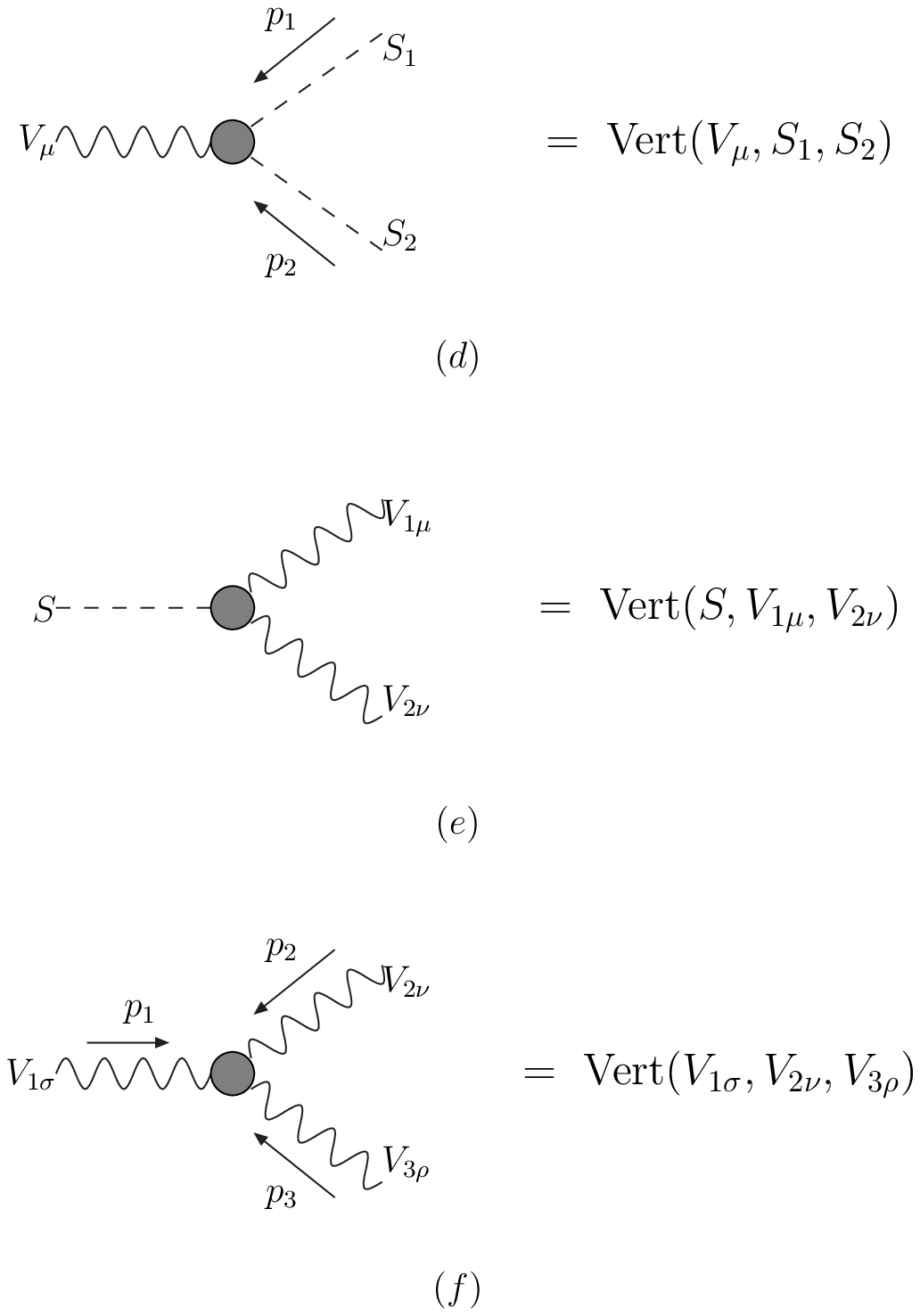}
\includegraphics[width=7cm]{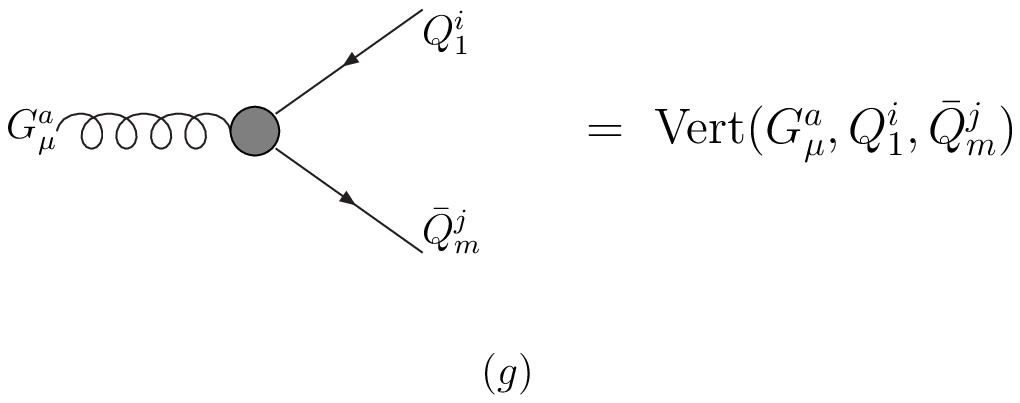}
\caption{\label{fig:ew3} All possible non-vanishing 3-point vertices
in EW.}
\end{figure}
\end{center}

All possible non-vanishing 3-point vertices in EW are shown in
Fig.\ref{fig:ew3}.

\begin{center}
\textbf{{\small Scalar-Fermion-Fermion vertices}}
 \vspace{0.1cm}
\end{center}

The generic effective vertex is shown in Fig.\ref{fig:ew3}~$(a)$
with the following expression
\begin{center}
\begin{eqnarray}
{\rm Vert}\left(S,F_1,\bar{F}_2\right)&=& \frac{e^3}{
\pi^2}~\left(C_{-}^{SF_1\bar{F}_2}\Omega^{-}+C^{SF_1\bar{F}_2}_{+}\Omega^{+}\right),
\end{eqnarray}
\end{center}
With its actual values of $S$, $F_1$, $\bar{F}_2$,
$C_{-}^{SF_1\bar{F}_2}$ and $C^{SF_1\bar{F}_2}_{+}$
\begin{center}
\begin{eqnarray*}
C_{-}^{HU_l\bar{U}_m}&=&\delta_{lm}~\frac{im_{U_l}}{8m_Ws_w}\left[\frac{\left(5+3g5s+8\lambda_{HV}\right)Q_{U_l}^2}
{16c_w^2}+\frac{1+g5s}{2}~\frac{1}{16s_w^2}\right.\nonumber\\&&
\sum_{g}\left(V_{U_l,D_g}V^{\dagger}_{D_g,U_l}\right)+\frac{1+g5s}{2}~\frac{I_{3U_l}}{c_w^2}\left(
\frac{I_{3U_l}}{8s_w^2}-\frac{\left(1+\lambda_{HV}\right)Q_{U_l}}{2}\right)\nonumber\\&&
\left.+\frac{1}{16m_W^2s_w^2}\sum_{g}\left(m_{D_g}^2V_{U_l,D_g}V^{\dagger}_{D_g,U_l}\right)\right],\nonumber\\
C_{+}^{HU_l\bar{U}_m}&=&C_{-}^{HU_l\bar{U}_m},\nonumber\\
C_{-}^{HD_l\bar{D}_m}&=&\delta_{lm}~\frac{im_{D_l}}{8m_Ws_w}\left[\frac{\left(1+3g5s+4\lambda_{HV}\right)Q_{D_l}^2}
{8c_w^2}+\frac{1+g5s}{2}~\frac{1}{16s_w^2}\right.\nonumber\\&&
\sum_{g}\left(V_{U_g,D_l}V^{\dagger}_{D_l,U_g}\right)+\frac{1+g5s}{2}~\frac{I_{3D_l}}{c_w^2}\left(
\frac{I_{3D_l}}{8s_w^2}-\frac{\left(1+\lambda_{HV}\right)Q_{D_l}}{2}\right)\nonumber\\&&
\left.+\frac{1}{16m_W^2s_w^2}\sum_{g}\left(m_{U_g}^2V_{U_g,D_l}V^{\dagger}_{D_l,U_g}\right)\right],\nonumber\\
C_{+}^{HD_l\bar{D}_m}&=&C_{-}^{HD_l\bar{D}_m},\nonumber\\
C_{-}^{HL_l\bar{L}_m}&=&\delta_{lm}~\frac{im_{L_l}}{8m_Ws_w}\left[\frac{\left(3+g5s+4\lambda_{HV}\right)Q_{L_l}^2}
{8c_w^2}+\frac{1+g5s}{2}~\frac{1}{16s_w^2}\right.\nonumber\\&&\left.
+\frac{1+g5s}{2}~\frac{I_{3L_l}}{c_w^2}\left(
\frac{I_{3L_l}}{8s_w^2}-\frac{\left(1+\lambda_{HV}\right)Q_{L_l}}{2}\right)\right],\nonumber\\
C_{+}^{HL_l\bar{L}_m}&=&C_{-}^{HL_l\bar{L}_m},\nonumber\\
C_{-}^{\phi^0U_l\bar{U}_m}&=&-g5s~\delta_{lm}~\frac{m_{U_l}}{4m_Ws_w}\left[\frac{\left(5+3g5s+8\lambda_{HV}\right)Q_{U_l}^2I_{3U_l}}
{16c_w^2}+\frac{1+g5s}{2}~\frac{1}{32s_w^2}\right.\nonumber\\&&
\sum_{g}\left(V_{U_l,D_g}V^{\dagger}_{D_g,U_l}\right)+\frac{1+g5s}{2}~\frac{I_{3U_l}}{c_w^2}\left(
\frac{1}{32s_w^2}-\frac{\left(1+\lambda_{HV}\right)Q_{U_l}I_{3U_l}}{2}\right)\nonumber\\&&
\left.-\frac{1}{16m_W^2s_w^2}\sum_{g}\left(m_{D_g}^2I_{3D_g}V_{U_l,D_g}V^{\dagger}_{D_g,U_l}\right)\right],\nonumber\\
C_{+}^{\phi^0U_l\bar{U}_m}&=&-C_{-}^{\phi^0U_l\bar{U}_m},
\end{eqnarray*}
\end{center}
\begin{center}
\begin{eqnarray}
C_{-}^{\phi^0D_l\bar{D}_m}&=&-g5s~\delta_{lm}~\frac{m_{D_l}}{4m_Ws_w}\left[\frac{\left(1+3g5s+4\lambda_{HV}\right)Q_{D_l}^2I_{3D_l}}
{8c_w^2}-\frac{1+g5s}{2}~\frac{1}{32s_w^2}\right.\nonumber\\&&
\sum_{g}\left(V_{U_g,D_l}V^{\dagger}_{D_l,U_g}\right)+\frac{1+g5s}{2}~\frac{I_{3D_l}}{c_w^2}\left(
\frac{1}{32s_w^2}-\frac{\left(1+\lambda_{HV}\right)Q_{D_l}I_{3D_l}}{2}\right)\nonumber\\&&
\left.-\frac{1}{16m_W^2s_w^2}\sum_{g}\left(m_{U_g}^2I_{3U_g}V_{U_g,D_l}V^{\dagger}_{D_l,U_g}\right)\right],\nonumber\\
C_{+}^{\phi^0D_l\bar{D}_m}&=&-C_{-}^{\phi^0D_l\bar{D}_m},\nonumber\\
C_{-}^{\phi^0L_l\bar{L}_m}&=&-g5s~\delta_{lm}~\frac{m_{L_l}}{4m_Ws_w}\left[\frac{\left(3+g5s+4\lambda_{HV}\right)Q_{L_l}^2I_{3L_1}}
{8c_w^2}-\frac{1+g5s}{2}~\frac{1}{32s_w^2}\right.\nonumber\\&&\left.
+\frac{1+g5s}{2}~\frac{I_{3L_l}}{c_w^2}\left(
\frac{1}{32s_w^2}-\frac{\left(1+\lambda_{HV}\right)Q_{L_l}I_{3L_l}}{2}\right)\right],\nonumber\\
C_{+}^{\phi^0L_l\bar{L}_m}&=&-C_{-}^{\phi^0L_l\bar{L}_m},\nonumber\\
C_{-}^{\phi^-U_l\bar{D}_m}&=&-\frac{1+g5s}{2}~\frac{im_{D_m}V^{\dagger}_{D_m,U_l}}{4\sqrt{2}m_Ws_w}
\left[-\frac{1}{16c_w^2}\left(1+8\left(1+\lambda_{HV}\right)Q_{U_l}Q_{D_m}\right)-\frac{3}{32s_w^2}\right.\nonumber\\&&
\left.-\frac{m_{U_l}^2}{16m_W^2s_w^2}+\frac{I_{3U_l}}{16c_w^2}\left(1+8\left(1+\lambda_{HV}\right)Q_{D_m}\right)\right]
+\frac{1-g5s}{2}~\frac{im_{U_l}V^{\dagger}_{D_m,U_l}}{4\sqrt{2}m_Ws_w}\nonumber\\&&
\left[-\frac{1}{16c_w^2}\left(1+8\left(1+\lambda_{HV}\right)Q_{U_l}Q_{D_m}\right)-\frac{m_{D_m}^2}{16m_W^2s_w^2}
+\frac{3m_{D_m}-m_{U_l}}{48m_{U_l}c_w^2}\right],\nonumber\\
C_{+}^{\phi^-U_l\bar{D}_m}&=&\frac{1+g5s}{2}~\frac{im_{U_l}V^{\dagger}_{D_m,U_l}}{4\sqrt{2}m_Ws_w}
\left[-\frac{1}{16c_w^2}\left(1+8\left(1+\lambda_{HV}\right)Q_{U_l}Q_{D_m}\right)-\frac{3}{32s_w^2}\right.\nonumber\\&&
\left.-\frac{m_{D_m}^2}{16m_W^2s_w^2}+\frac{I_{3D_m}}{16c_w^2}\left(1+8\left(-1+\lambda_{HV}\right)Q_{U_l}\right)\right]
-\frac{1-g5s}{2}~\frac{im_{D_m}V^{\dagger}_{D_m,U_l}}{4\sqrt{2}m_Ws_w}\nonumber\\&&
\left[-\frac{1}{16c_w^2}\left(1+8\left(1+\lambda_{HV}\right)Q_{U_l}Q_{D_m}\right)-\frac{m_{U_l}^2}{16m_W^2s_w^2}
-\frac{5m_{D_m}-3m_{U_l}}{48m_{D_m}c_w^2}\right],\nonumber\\
C_{\pm}^{\phi^+D_l\bar{U}_m}&=&-\left(C_{\mp}^{\phi^-U_m\bar{D}_l}\right)^*,\nonumber\\
C_{-}^{\phi^-\nu_l\bar{e}_m}&=&-\delta_{lm}~\frac{im_{e_l}}{4\sqrt{2}m_Ws_w}
\left[\frac{Q_{e_l}}{16c_w^2}-\frac{1+g5s}{2}~\frac{3}{32s_w^2}\right.\nonumber\\&&\left.
+\frac{1+g5s}{2}~\frac{I_{3\nu_l}}{16c_w^2}\left(1+8\left(1+\lambda_{HV}\right)Q_{e_l}\right)\right],\nonumber\\
C_{+}^{\phi^-\nu_l\bar{e}_m}&=&0,\nonumber\\
C_{\pm}^{\phi^+e_l\bar{\nu}_m}&=&-\left(C_{\mp}^{\phi^-\nu_m\bar{e}_l}\right)^*.
\end{eqnarray}
\end{center}
\vspace{0.5cm}

\begin{center}
\textbf{{\small Vector-Fermion-Fermion vertices}}
 \vspace{0.1cm}
\end{center}

The generic effective vertex is shown in Fig.\ref{fig:ew3}~$(b)$
with the following expression
\begin{center}
\begin{eqnarray}
{\rm Vert}\left(V_{\mu},F_1,\bar{F}_2\right)&=& \frac{i e^3}{
\pi^2}~\left(C_{-}^{VF_1\bar{F}_2}\Omega^-+C_{+}^{VF_1\bar{F}_2}\Omega^+\right)~\gamma_{\mu},
\end{eqnarray}
\end{center}
with the actual values of $V$, $F_1$, $\bar{F}_2$,
$C_{-}^{VF_1\bar{F}_2}$ and $C_{+}^{VF_1\bar{F}_2}$
\begin{center}
\begin{eqnarray*}
C_{-}^{AU_l\bar{U}_m}&=&\delta_{lm}~\frac{1}{4}\left[\frac{1+\lambda_{HV}}{4c_w^2}Q_{U_l}^3+\frac{m_{U_l}^2}{8s_w^2m_W^2}
\left(\frac{1}{2}\sum_{g}\left(V_{U_l,D_g}V^{\dagger}_{D_g,U_l}Q_{D_g}\right)\right.\right.\nonumber\\&&
\left.\left.+\frac{Q_{U_l}}{4}+Q_{U_l}I_{3U_l}^2\right)\right],\nonumber\\
C_{+}^{AU_l\bar{U}_m}&=&\delta_{lm}~\frac{1}{4}\left\{\frac{1+\lambda_{HV}}{4c_w^2}Q_{U_l}^3-\frac{2+\left(1+g5s\right)
~\lambda_{HV}}{4c_w^2}Q_{U_l}^2I_{3U_l}\right.\nonumber\\&&
+\frac{2+\left(1+g5s\right)\lambda_{HV}}{8s_w^2c_w^2}Q_{U_l}I_{3U_l}^2+\frac{1}{4s_w^2}\left[\frac{1}{4m_W^2}
\sum_{g}\left(V_{U_l,D_g}V^{\dagger}_{D_g,U_l}m_{D_g}^2Q_{D_g}\right)\right.\nonumber\\&&
+\frac{1+4I_{3U_l}^2}{8m_W^2}m_{U_l}^2Q_{U_l}+\frac{2+\left(1+g5s\right)\lambda_{HV}}{4}\nonumber\\&&
\left.\left.\sum_{g}\left(V_{U_l,D_g}V^{\dagger}_{D_g,U_l}
\left(\frac{1+g5s}{2}+Q_{D_g}\right)\right)\right]\right\},\nonumber\\
C_{-}^{AD_l\bar{D}_m}&=&\delta_{lm}~\frac{1}{4}\left[\frac{1+\lambda_{HV}}{4c_w^2}Q_{D_l}^3+\frac{m_{D_l}^2}{8s_w^2m_W^2}
\left(\frac{1}{2}\sum_{g}\left(V_{U_g,D_l}V^{\dagger}_{D_l,U_g}Q_{U_g}\right)\right.\right.\nonumber\\&&
\left.\left.+\frac{Q_{D_l}}{4}+Q_{D_l}I_{3D_l}^2\right)\right],\nonumber\\
C_{+}^{AD_l\bar{D}_m}&=&\delta_{lm}~\frac{1}{4}\left\{\frac{1+\lambda_{HV}}{4c_w^2}Q_{D_l}^3-\frac{2+\left(1+g5s\right)
~\lambda_{HV}}{4c_w^2}Q_{D_l}^2I_{3D_l}\right.\nonumber\\&&
+\frac{2+\left(1+g5s\right)\lambda_{HV}}{8s_w^2c_w^2}Q_{D_l}I_{3D_l}^2+\frac{1}{4s_w^2}\left[\frac{1}{4m_W^2}
\sum_{g}\left(V_{U_g,D_l}V^{\dagger}_{D_l,U_g}m_{U_g}^2Q_{U_g}\right)\right.\nonumber\\&&
+\frac{1+4I_{3D_l}^2}{8m_W^2}m_{D_l}^2Q_{D_l}+\frac{2+\left(1+g5s\right)\lambda_{HV}}{4}\nonumber\\&&
\left.\left.\sum_{g}\left(V_{U_l,D_g}V^{\dagger}_{D_g,U_l}
\left(Q_{U_g}-\frac{1+g5s}{2}\right)\right)\right]\right\},\nonumber\\
C_{-}^{Ae_l\bar{e}_m}&=&\delta_{lm}~\frac{1}{4}\left[\frac{1+\lambda_{HV}}{4c_w^2}Q_{e_l}^3+\frac{m_{e_l}^2}{8m_W^2s_w^2}
\left(\frac{Q_{e_l}}{4}+Q_{e_l}I_{3e_l}^2\right)\right],
\end{eqnarray*}
\end{center}
\begin{center}
\begin{eqnarray*}
C_{+}^{Ae_l\bar{e}_m}&=&\delta_{lm}~\frac{1}{4}\left[\frac{1+\lambda_{HV}}{4c_w^2}Q_{e_l}^3-\frac{2+
\left(1+g5s\right)\lambda_{HV}}{4c_w^2}Q_{e_l}^2I_{3e_l}\right.\nonumber\\&&+\frac{2+\left(1+g5s\right)\lambda_{HV}}{8s_w^2c_w^2}
Q_{e_l}I_{3e_l}^2+\frac{1}{4s_w^2}\left(\frac{1+4I_{3e_l}^2}{8m_W^2}m_{e_l}^2Q_{e_l}\right.\nonumber\\&&
\left.\left.-\frac{1+g5s}{2}~\frac{1+\lambda_{HV}}{2}\right)\right],\nonumber\\
C_{-}^{A\nu_l\bar{\nu}_m}&=&0,\nonumber\\
C_{+}^{A\nu_l\bar{\nu}_m}&=&\delta_{lm}~\frac{1}{32s_w^2}\left[\frac{m_{e_l}^2Q_{e_l}}{2m_W^2}+\left(Q_{e_l}
+\frac{1+g5s}{2}\right)\left(\frac{1+g5s}{2}\lambda_{HV}+1\right)\right],\nonumber\\
C_{-}^{ZU_l\bar{U}_m}&=&\delta_{lm}~\frac{1}{8c_w}\left\{\frac{\left(1+\lambda_{HV}\right)s_w}{2c_w^2}Q_{U_l}^3+\frac{m_{U_l}^2}{8s_wm_W^2}
\left[\sum_{g}\left(V_{U_l,D_g}V^{\dagger}_{D_g,U_l}\right.\right.\right.\nonumber\\&&\left.\left.\left(Q_{D_g}-\frac{1+g5s}{2}~\frac{I_{3D_g}}{s_w^2}\right)\right)
+Q_{U_l}-\frac{1+g5s}{2}~\frac{I_{3U_l}}{s_w^2}\right]\nonumber\\&&
\left.-\frac{1-g5s}{2}\frac{1+\lambda_{HV}}{2s_wc_w^2}Q_{U_l}^2I_{3U_l}\right\},\nonumber\\
C_{+}^{ZU_l\bar{U}_m}&=&\delta_{lm}~\frac{1}{8c_w}\left\{\frac{\left(1+\lambda_{HV}\right)s_w}{2c_w^2}Q_{U_l}^3
-\frac{\left(2+\left(1+g5s\right)
~\lambda_{HV}\right)\left(4s_w^2+1+g5s\right)}{8c_w^2s_w}\right.\nonumber\\&&Q_{U_l}^2I_{3U_l}
+\left(2+g5s\right)\frac{2+\left(1+g5s\right)\lambda_{HV}}{4s_wc_w^2}Q_{U_l}I_{3U_l}^2\nonumber\\&&-\frac{1+g5s}{2}~
\frac{\left(1+\lambda_{HV}\right)I_{3U_l}^3}{2s_w^3c_w^2}-\frac{1-g5s}{2}~\frac{m_{U_l}^2I_{3U_l}}{8m_W^2s_w^3}
+\frac{1}{2s_w}\left[\frac{1}{4m_W^2}\right. \nonumber\\&&
\sum_{g}\left(V_{U_l,D_g}V^{\dagger}_{D_g,U_l}m_{D_g}^2\left(Q_{D_g}-\frac{1-g5s}{2}~\frac{I_{3D_g}}{s_w^2}\right)\right)
\nonumber\\&&
+\frac{1+4I_{3U_l}^2}{8m_W^2}m_{U_l}^2Q_{U_l}+\frac{2+\left(1+g5s\right)\lambda_{HV}}{4}\nonumber\\&&
\left.\left.\sum_{g}\left(V_{U_l,D_g}V^{\dagger}_{D_g,U_l}
\left(Q_{D_g}-\frac{1+g5s}{2}~\frac{c_w^2+I_{3D_g}}{s_w^2}\right)\right)\right]\right\},\nonumber\\
C_{-}^{ZD_l\bar{D}_m}&=&\delta_{lm}~\frac{1}{8c_w}\left\{\frac{\left(1+\lambda_{HV}\right)s_w}{2c_w^2}Q_{D_l}^3+\frac{m_{D_l}^2}{8s_wm_W^2}
\left[\sum_{g}\left(V_{U_g,D_l}V^{\dagger}_{D_l,U_g}\right.\right.\right.\nonumber\\&&\left.\left.\left(Q_{U_g}-\frac{1+g5s}{2}~\frac{I_{3U_g}}{s_w^2}\right)\right)
+Q_{D_l}-\frac{1+g5s}{2}~\frac{I_{3D_l}}{s_w^2}\right]\nonumber\\&&
\left.-\frac{1-g5s}{2}\frac{1+\lambda_{HV}}{2s_wc_w^2}Q_{D_l}^2I_{3D_l}\right\},
\end{eqnarray*}
\end{center}
\begin{center}
\begin{eqnarray*}
C_{+}^{ZD_l\bar{D}_m}&=&\delta_{lm}~\frac{1}{8c_w}\left\{\frac{\left(1+\lambda_{HV}\right)s_w}{2c_w^2}Q_{D_l}^3
-\frac{\left(2+\left(1+g5s\right)
~\lambda_{HV}\right)\left(4s_w^2+1+g5s\right)}{8c_w^2s_w}\right.\nonumber\\&&Q_{D_l}^2I_{3D_l}
+\left(2+g5s\right)\frac{2+\left(1+g5s\right)\lambda_{HV}}{4s_wc_w^2}Q_{D_l}I_{3D_l}^2\nonumber\\&&-\frac{1+g5s}{2}~
\frac{\left(1+\lambda_{HV}\right)I_{3D_l}^3}{2s_w^3c_w^2}-\frac{1-g5s}{2}~\frac{m_{D_l}^2I_{3D_l}}{8m_W^2s_w^3}
+\frac{1}{2s_w}\left[\frac{1}{4m_W^2}\right. \nonumber\\&&
\sum_{g}\left(V_{U_g,D_l}V^{\dagger}_{D_l,U_g}m_{U_g}^2\left(Q_{U_g}-\frac{1-g5s}{2}~\frac{I_{3U_g}}{s_w^2}\right)\right)
\nonumber\\&&
+\frac{1+4I_{3D_l}^2}{8m_W^2}m_{D_l}^2Q_{D_l}+\frac{2+\left(1+g5s\right)\lambda_{HV}}{4}\nonumber\\&&
\left.\left.\sum_{g}\left(V_{U_g,D_l}V^{\dagger}_{D_l,U_g}
\left(Q_{U_g}+\frac{1+g5s}{2}~\frac{c_w^2-I_{3U_g}}{s_w^2}\right)\right)\right]\right\},\nonumber\\
C_{-}^{Ze_l\bar{e}_m}&=&\delta_{lm}~\frac{1}{8c_w}\left\{\frac{\left(1+\lambda_{HV}\right)s_w}{2c_w^2}Q_{e_l}^3
+\frac{m_{e_l}^2}{4s_wm_W^2}
\left[-\frac{1+g5s}{2}~\frac{1}{4s_w^2}+\frac{1}{2}\right.\right.\nonumber\\&&\left.\left.\left(Q_{e_l}-\frac{1+g5s}{2}~\frac{I_{3e_l}}{s_w^2}\right)\right]
-\frac{1-g5s}{2}\frac{1+\lambda_{HV}}{2s_wc_w^2}Q_{e_l}^2I_{3e_l}\right\},\nonumber\\
C_{+}^{Ze_l\bar{e}_m}&=&\delta_{lm}~\frac{1}{16c_w}\left\{\frac{\left(1+\lambda_{HV}\right)s_w}{c_w^2}Q_{e_l}^3
-\frac{\left(2+\left(1+g5s\right)
~\lambda_{HV}\right)\left(4s_w^2+1+g5s\right)}{4c_w^2s_w}\right.\nonumber\\&&Q_{e_l}^2I_{3e_l}
+\left(2+g5s\right)\frac{2+\left(1+g5s\right)\lambda_{HV}}{2s_wc_w^2}Q_{e_l}I_{3e_l}^2\nonumber\\&&-\frac{1+g5s}{2}~
\frac{\left(1+\lambda_{HV}\right)I_{3e_l}^3}{s_w^3c_w^2}-\frac{1-g5s}{2}~\frac{m_{e_l}^2I_{3e_l}}{4m_W^2s_w^3}
\nonumber\\&&\left.+\frac{1}{2s_w}\left[\frac{m_{e_l}^2Q_{e_l}\left(1+4I_{3e_l}^2\right)}{4m_W^2}+
\frac{1+g5s}{2}~\frac{\left(1+\lambda_{HV}\right)
\left(c_w^2-I_{3\nu_l}\right)}{s_w^2}
\right]\right\},\nonumber\\
C_{-}^{Z\nu_l\bar{\nu}_m}&=&0,\nonumber\\
C_{+}^{Z\nu_l\bar{\nu}_m}&=&\delta_{lm}~\frac{1}{16c_w}\left\{-\frac{1+g5s}{2}~\frac{\left(1+\lambda_{HV}\right)
I_{3\nu_l}^3}{s_w^3c_w^2}+\frac{1}{2s_w}\left[\frac{m_{e_l}^2Q_{e_l}}{2m_W^2}\right.\right.\nonumber\\&&
\left.+\left(1+\frac{1+g5s}{2}\lambda_{HV}\right)\left(Q_{e_l}-\frac{1+g5s}{2}~\frac{c_w^2+I_{3e_l}}{s_w^2}\right)\right]
\nonumber\\&&\left.-\frac{1-g5s}{2}~\frac{I_{3e_l}m_{e_l}^2}{4m_W^2s_w^3}\right\},
\end{eqnarray*}
\end{center}
\begin{center}
\begin{eqnarray}
C_{-}^{W^-U_l\bar{D}_m}&=&-\frac{1-g5s}{2}~\frac{V^{\dagger}_{Dm,U_l}}{16\sqrt{2}s_w}\left(1+\lambda_{HV}\right)
\frac{Q_{U_l}Q_{D_m}}{c_w^2},\nonumber\\
C_{+}^{W^-U_l\bar{D}_m}&=&\frac{1+g5s}{2}~\frac{V^{\dagger}_{Dm,U_l}}{16\sqrt{2}s_w}\left(1+\lambda_{HV}\right)
\left[\frac{Q_{D_m}I_{3U_l}+Q_{U_l}I_{3D_m}-Q_{U_l}Q_{D_m}}{c_w^2}\right.\nonumber\\&&
\left.-\frac{1}{s_w^2}+\frac{1}{4s_w^2c_w^2}\right],\nonumber\\
C_{\pm}^{W^+D_l\bar{U}_m}&=&\left(C_{\pm}^{W^-U_m\bar{D}_l}\right)^*,\nonumber\\
C_{-}^{W^-\nu_l\bar{e}_m}&=&0,\nonumber\\
C_{+}^{W^-\nu_l\bar{e}_m}&=&\frac{1+g5s}{2}~\delta_{lm}~\frac{1}{16\sqrt{2}s_w}\left[
\frac{Q_{e_l}I_{3\nu_l}}{c_w^2}-\frac{1}{s_w^2}+\frac{1}{4s_w^2c_w^2}\right]\left(1+\lambda_{HV}\right),\nonumber\\
C_{\pm}^{W^+e_l\bar{\nu}_m}&=&\left(C_{\pm}^{W^-\nu_m\bar{e}_l}\right)^*.
\end{eqnarray}
\end{center}
 \vspace{0.5cm}

\begin{center}
\textbf{{\small Scalar-Scalar-Scalar vertices}}
 \vspace{0.1cm}
\end{center}

The generic effective vertex is shown in Fig.\ref{fig:ew3}~$(c)$
with the following expression
\begin{center}
\begin{eqnarray}
{\rm Vert}\left(S_{1},S_{2},S_3\right)&=& \frac{i e^3}{
\pi^2}~C^{S_1S_2S_3},
\end{eqnarray}
\end{center}
with its actual values of $S_1$, $S_2$, $S_3$ and $C^{S_1S_2S_3}$
\begin{center}
\begin{eqnarray*}
C^{HH\phi^0}&=&0,~~~~C^{\phi^0\phi^0\phi^0}=0,~~~~C^{\phi^0\phi^-\phi^+}=0,\nonumber\\
C^{HHH}&=&\frac{3}{32s_w^2}\left[\frac{1-4\lambda_{HV}}{2}m_W+\frac{1}{m_W^3}\left(\sum_{l}m_{e_l}^4
+N_c\sum_{l}m_{Q_l}^4\right)\right.\nonumber\\&&
\left.+\left(1+\frac{1}{2c_w^2}\right)\frac{m_{H}^2}{4m_W}+\frac{\left(1-4\lambda_{HV}\right)m_W}{4c_w^4}\right],\nonumber\\
C^{H\phi^0\phi^0}&=&\frac{1}{8s_w^3}\left[\frac{1-4\lambda_{HV}}{8}m_W+\frac{2g5s-1}{4m_W^3}\left(\sum_{l}m_{e_l}^4
+N_c\sum_{l}m_{Q_l}^4\right)\right.\nonumber\\&&\left.+\left(1+\frac{1}{2c_w^2}\right)\frac{m_{H}^2}{4m_W}
+\frac{\left(1-4\lambda_{HV}\right)m_W}{16c_w^4}\right],
\end{eqnarray*}
\end{center}
\begin{center}
\begin{eqnarray}
C^{H\phi^-\phi^+}&=&\frac{1}{32s_w^3}\left\{\frac{1-4\lambda_{HV}}{4}m_W
\left(3+\frac{s_w^2\left(1+c_w^2\right)}{c_w^4}\right)+\frac{\left(1+2c_w^2\right)
m_H^2}{8c_w^2m_W}\right.\nonumber\\&&\frac{1}{m_W^3}\left[\frac{1+g5s}{2}~
\sum_{l}m_{e_l}^4+N_c\sum_{l,m}\left[V_{U_l,D_m}V^{\dagger}_{D_m,U_l}\right.\right.\nonumber\\&&
\left(\frac{1+g5s}{2}~\left(m_{U_l}^4+m_{D_m}^4\right)-\frac{1-g5s}{2}~2m_{U_l}m_{D_m}
\right.\nonumber\\&&\left.\left.\left.\left.\left(m_{U_l}^2
+m_{U_l}m_{D_m}+m_{D_m}^2\right)\right)\right]\right]\right\}.
\end{eqnarray}
\end{center}
\vspace{0.5cm}

\begin{center}
\textbf{{\small Vector-Scalar-Scalar vertices}}
 \vspace{0.1cm}
\end{center}

The generic effective vertex is shown in Fig.\ref{fig:ew3}~$(d)$
with the following expression
\begin{center}
\begin{eqnarray}
{\rm Vert}\left(V_{\mu},S_1,S_2\right)&=&
\frac{e^3}{\pi^2}~C^{VS_1S_2}_{\mu},
\end{eqnarray}
\end{center}
with the actual values of $V$, $S_1$, $S_2$ and $C^{VS_1S_2}_{\mu}$
\begin{center}
\begin{eqnarray*}
C^{AHH}_{\mu}&=&0,~~~C^{ZHH}_{\mu}=0,~~~C^{A\phi^0\phi^0}_{\mu}=0,~~~C^{Z\phi^0\phi^0}_{\mu}=0,\nonumber\\
C^{A\phi^0H}_{\mu}&=&\frac{5}{192s_w^2}~\left(p_1-p_2\right)_{\mu},\nonumber\\
C^{Z\phi^0H}_{\mu}&=&-\frac{1}{96s_wc_w}~\left[\frac{1+2c_w^2+20c_w^4}{8s_w^2c_w^2}+\frac{1}{s_w^2m_W^2}
\left(\sum_{l}m_{e_l}^2+N_c\sum_{l}m_{Q_l}^2\right)\right]\nonumber\\&&
~\left(p_1-p_2\right)_{\mu}-\frac{1-g5s}{2}~\frac{3}{96s_wc_w}\left[\frac{1}{s_w^2m_W^2}\left(\sum_{l}m_{e_l}^2+N_c\sum_{l}m_{Q_l}^2
\right)\right]\left(p_{2}\right)_{\mu},\nonumber\\
C^{A\phi^+\phi^-}_{\mu}&=&g5s~\frac{i}{48s_w^2}~\left[\frac{1+12c_w^2}{8c_w^2}+\frac{3+g5s}{4m_W^2}
\left(\sum_{l}\left(m_{e_l}^2\right)+N_c\sum_{l,m}\left(V_{U_l,D_m}V^{\dagger}_{D_m,U_l}
\right.\right.\right.\nonumber\\&&\left.\left.\left.\left(m_{U_l}^2+m_{D_m}^2-\frac{1-g5s}{2}
~2m_{U_l}m_{D_m}\right)\right)\right)\right]~\left(p_1-p_2\right)_{\mu},\nonumber\\
C^{Z\phi^+\phi^-}_{\mu}&=&g5s~\frac{i}{48s_wc_w}~\left\{\frac{1-24c_w^4}{16c_w^2s_w^2}+\frac{3+g5s}{4m_W^2}
\left[\sum_{l}\left(m_{e_l}^2\left(1-\frac{1}{2s_w^2}\right)\right)
\right.\right.\nonumber\\&&+N_c\sum_{l,m}\left[V_{U_l,D_m}V^{\dagger}_{D_m,U_l}
\left(m_{U_l}^2+m_{D_m}^2-\frac{1-g5s}{2}
~2m_{U_l}m_{D_m}\right)\right.\nonumber\\&&
\left.\left.\left.\left(1-\frac{1}{2s_w^2}\right)\right]\right]\right\}~\left(p_1-p_2\right)_{\mu},
\end{eqnarray*}
\end{center}
\begin{center}
\begin{eqnarray}
C^{W^-H\phi^+}_{\mu}&=&\frac{i}{96s_w^3}\left\{\frac{1+22c_w^2}{8c_w^2}+\frac{1}{m_W^2}\left[\sum_{l}m_{e_l}^2
+N_c\sum_{l,m}\left(V_{U_l,D_m}V^{\dagger}_{D_m,U_l}\right.\right.\right.\nonumber\\&&\left.\left.\left.
\left(m_{U_l}^2+m_{D_m}^2\right)\right)\right]\right\}~\left(p_1-p_2\right)_{\mu}-\frac{1-g5s}{2}~\frac{i}{96s_w^3}\frac{3}{2m_W^2}
\nonumber\\&&\left[\sum_{l}m_{e_l}^2
+N_c\sum_{l,m}\left(V_{U_l,D_m}V^{\dagger}_{D_m,U_l}
\left(m_{U_l}+m_{D_m}\right)^2\right)\right]~\left(p_1\right)_{\mu},\nonumber\\
C^{W^+\phi^-H}_{\mu}&=&\frac{i}{96s_w^3}\left\{\frac{1+22c_w^2}{8c_w^2}+\frac{1}{m_W^2}\left[\sum_{l}m_{e_l}^2
+N_c\sum_{l,m}\left(V_{U_l,D_m}V^{\dagger}_{D_m,U_l}\right.\right.\right.\nonumber\\&&\left.\left.\left.
\left(m_{U_l}^2+m_{D_m}^2\right)\right)\right]\right\}~\left(p_1-p_2\right)_{\mu}+\frac{1-g5s}{2}~\frac{i}{96s_w^3}\frac{3}{2m_W^2}
\nonumber\\&&\left[\sum_{l}m_{e_l}^2
+N_c\sum_{l,m}\left(V_{U_l,D_m}V^{\dagger}_{D_m,U_l}
\left(m_{U_l}+m_{D_m}\right)^2\right)\right]~\left(p_2\right)_{\mu},\nonumber\\
C^{W^-\phi^+\phi^0}_{\mu}&=&\frac{1}{48s_w^3}\left\{-\frac{1+22c_w^2}{16c_w^2}+\frac{3+g5s}{4m_W^2}\left[
\sum_{l}\left(m_{e_l}^2I_{3e_l}\right)-\frac{N_c}{2}\sum_{l,m}\left(V_{U_l,D_m}V^{\dagger}_{D_m,U_l}\right.\right.\right.\nonumber\\&&
\left.\left.\left.\left(m_{U_l}^2+m_{D_m}^2-\left(1-g5s\right)m_{U_l}m_{D_m}\right)\right)\right]\right\}
~\left(p_1-p_2\right)_{\mu}\nonumber\\&&
-\frac{1-g5s}{2}~\frac{3}{96m_W^2s_w^3}
\left[\sum_{l}\left(m_{e_l}^2I_{3e_l}\right)\right.\nonumber\\&&\left.-\frac{N_c}{2}\sum_{l,m}\left(V_{U_l,D_m}V^{\dagger}_{D_m,U_l}
\left(m_{U_l}-m_{D_m}\right)^2\right)\right]\left(p_1\right)_{\mu},\nonumber\\
C^{W^+\phi^-\phi^0}_{\mu}&=&C^{W^-\phi^+\phi^0}_{\mu}.
\end{eqnarray}
\end{center}
\vspace{0.5cm}

\begin{center}
\textbf{{\small Scalar-Vector-Vector vertices}}
 \vspace{0.1cm}
\end{center}

The generic effective vertex is shown in Fig.\ref{fig:ew3}~$(e)$
with the following expression
\begin{center}
\begin{eqnarray}
{\rm Vert}\left(S,V_{1\mu},V_{2\nu}\right)&=& \frac{i e^3}{
\pi^2}~C^{SV_1V_2}~g_{\mu\nu},
\end{eqnarray}
\end{center}
with the actual values of $S$, $V_1$, $V_2$ and $C^{SV_1V_2}$
\begin{center}
\begin{eqnarray}
C^{\phi^0AA}&=&0,~~~C^{\phi^0AZ}=0,~~~C^{\phi^0ZZ}=0,~~~C^{\phi^0W^-W^+}=0,\nonumber\\
C^{HAA}&=&-\frac{1}{8s_w}\left[\frac{1}{m_W}\left(\sum_{l}\left(m_{L_l}^2Q_{L_l}^2\right)+N_c\sum_{l}
\left(m_{Q_l}^2Q_{Q_l}^2\right)\right)+\frac{m_W}{2}\right],\nonumber\\
C^{HAZ}&=&\frac{1}{8c_w}\left\{\frac{1}{m_W}\left[\sum_{l}\left(m_{L_l}^2Q_{L_l}\left(\frac{I_{3L_l}}{2s_w^2}
-Q_{L_l}\right)\right)\right.\right.\nonumber\\&&\left.\left.
+N_c\sum_{l}\left(m_{Q_l}^2Q_{Q_l}\left(\frac{I_{3Q_l}}{2s_w^2}-Q_{Q_l}\right)\right)\right]
+\frac{m_W^2\left(1+2c_w^2\right)}{4s_w^2}\right\},\nonumber\\
C^{HZZ}&=&\frac{1}{8}\left\{\frac{1}{m_Wc_w^2}\left[\sum_{l}\left(m_{L_l}^2\left(\frac{Q_{L_l}I_{3L_l}}
{s_w}-Q_{L_l}^2s_w-\frac{1+g5s}{2}~\frac{I_{3L_l}^2}{s_w^3}\right)\right)\right.\right.\nonumber\\&&
\left.+N_c\sum_{l}\left(m_{Q_l}^2\left(\frac{Q_{Q_l}I_{3L_l}}{s_w}-Q_{Q_l}^2s_w-\frac{1+g5s}{2}
~\frac{I_{3Q_l}^2}{s_w^3}\right)\right)\right]\nonumber\\&&
\left.+\frac{m_W\left(s_w^2-2\right)}{2s_w^3}\right\},\nonumber\\
C^{HW^-W^+}&=&-\frac{1}{8s_w^3}\left\{\frac{1+g5s}{2}~\frac{1}{4m_W}\left[\sum_{l}m_{L_l}^2\right.\right.\nonumber\\&&
\left.\left.+N_c\sum_{l,m}
\left(V_{U_l,D_m}V^{\dagger}_{D_m,U_l}\left(m_{U_l}^2+m_{D_m}^2\right)\right)\right]+m_W\right\},\nonumber\\
C^{\phi^-AW^+}&=&C^{\phi^+W^-A}=\frac{K}{32s_w^2},\nonumber\\
C^{\phi^-ZW^+}&=&C^{\phi^+W^-Z}=\frac{K}{32s_wc_w},
\end{eqnarray}
\end{center}
where
\begin{center}
\begin{eqnarray}
K=m_W+\frac{N_c}{m_W}\sum_{l,m}\left(V_{U_l,D_m}V^{\dagger}_{D_m,U_l}\left(Q_{U_l}m_{D_m}^2-Q_{D_m}m_{U_l}^2\right)\right).
\end{eqnarray}
\end{center}
\vspace{0.5cm}

\begin{center}
\textbf{{\small Vector-Vector-Vector vertices}}
 \vspace{0.1cm}
\end{center}

The generic effective vertex is shown in Fig.\ref{fig:ew3}~$(f)$
with the following expression
\begin{center}
\begin{eqnarray}
{\rm
Vert}\left(V_{1\mu},V_{2\nu},V_{3\rho}\right)&=&\frac{ie^3}{\pi^2}~C_{\mu\nu\rho}^{V_1V_2V_3},
\end{eqnarray}
\end{center}
with the actual values of $V_1$, $V_2$, $V_3$, and
$C_{\mu\nu\rho}^{V_1V_2V_3}$
\begin{center}
\begin{eqnarray}
C^{AAA}_{\mu\nu\rho}&=&C^{AAZ}_{\mu\nu\rho}=C^{AZZ}_{\mu\nu\rho}=C^{ZZZ}_{\mu\nu\rho}=0,\nonumber\\
C^{AW^+W^-}_{\mu\nu\rho}&=&K~
V_{1\mu\nu\rho}+\frac{1+3g5s}{2}~\frac{i}{288s_w^2}\left[\left(N_c\sum_{l,m=1}^3\left(V_{U_l,D_m}V^{\dagger}_{D_m,U_l}\right)-9\right)
\right.\nonumber\\&&\left.\left(\varepsilon_{\left(p_2-p_3\right)\mu\nu\rho}+\frac{1-g5s}{2}~3iV_{2\mu\nu\rho}\right)\right]
+\frac{1-g5s}{2}~\frac{1}{8s_w^2}V_{2\mu\nu\rho},
\nonumber\\
C^{ZW^+W^-}&=&-\frac{c_w}{s_w}\left(K-\frac{1-g5s}{2}~\frac{1}{4s_w^2}\right)V_{1\mu\nu\rho}+\frac{1-g5s}{2}
~\frac{1}{8c_ws_w}V_{3\mu\nu\rho}\nonumber\\&&+\frac{1+3g5s}{2}~\frac{1}{288c_ws_w^3}
\left(N_c\sum_{l,m=1}^3\left(V_{U_l,D_m}V^{\dagger}_{D_m,U_l}\right)-9\right)\nonumber\\&&
\left(is_w^2
\varepsilon_{\left(p_2-p_3\right)\mu\nu\rho}-\frac{1-g5s}{2}~3\left(-2V_{1\mu\nu\rho}
+s_w^2V_{2\mu\nu\rho}\right)\right),
\end{eqnarray}
\end{center}
where
\begin{center}
\begin{eqnarray}
K&=&\frac{7+4\lambda_{HV}}{96s_w^2}+\frac{1}{48s_w^2}\left(3+N_c\sum_{l,m=1}^3
\left(V_{U_l,D_m}V^{\dagger}_{D_m,U_l}\right)\right),\nonumber\\
V_{1\mu\nu\rho}&=&g_{\mu\nu}\left(p_2-p_1\right)_{\rho}+g_{\nu\rho}\left(p_3-p_2\right)_{\mu}+g_{\rho\nu}
\left(p_1-p_3\right)_{\nu},\nonumber\\
V_{2\mu\nu\rho}&=&3g_{\mu\nu}\left(p_1\right)_{\rho}+g_{\nu\rho}\left(p_2-p_3\right)_{\mu}-3g_{\rho\nu}
\left(p_1\right)_{\nu},\nonumber\\
V_{3\mu\nu\rho}&=&g_{\mu\nu}\left(p_2-p_3\right)_{\rho}+g_{\mu\rho}\left(p_2-p_3\right)_{\nu}-g_{\nu\rho}
\left(p_2-p_3\right)_{\mu}.
\end{eqnarray}
\end{center}
\vspace{0.5cm}

\begin{center}
\textbf{{\small Gluon-Quark-Quark vertices}}
 \vspace{0.1cm}
\end{center}

In previous subsection, we have listed all mixed $R$ QCD/EW vertices
with internal QCD particles. The remaining non-vanishing mixed $R$
QCD/EW vertices with EW internal particles are those
Gluon-Quark-Quark vertices. Its generic diagram is shown in
Fig.\ref{fig:ew3} $(g)$ with the following expression
\begin{center}
\begin{eqnarray}
{\rm
Vert}\left(G^a_{\mu},Q^i_l,\bar{Q}^j_m\right)&=&\frac{ig_se^2}{\pi^2}~T^a_{ji}~
\left(C_{-}^{Q_l\bar{Q}_m}\Omega^-+C_{+}^{Q_l\bar{Q}_m}\Omega^+\right)\gamma_{\mu}.
\end{eqnarray}
\end{center}
The actual values of $Q_l$, $\bar{Q}_m$, $C_{-}^{Q_l\bar{Q}_m}$ and
$C_{+}^{Q_l\bar{Q}_m}$ are
\begin{center}
\begin{eqnarray}
C^{U_l\bar{U}_m}_{-}&=&\delta_{lm}~\frac{1}{16}\left[\left(1+\lambda_{HV}\right)\frac{Q_{U_l}^2}{c_w^2}
+\frac{m_{U_l}^2}{2s_w^2m_W^2}\left(\frac{1}{2}\sum_{g}\left(V_{U_l,D_g}V^{\dagger}_{D_g,U_l}\right)\right.\right.\nonumber\\&&
\left.\left.+\frac{1}{4}+I_{3U_l}^2\right)\right],\nonumber\\
C^{U_l\bar{U}_m}_{+}&=&\delta_{lm}~\frac{1}{16}\left[\left(1+\lambda_{HV}\right)\frac{Q_{U_l}^2}{c_w^2}+\frac{2+
\left(1+g5s\right)\lambda_{HV}}{2c_w^2}\left(\frac{I_{3U_l}^2}{s_w^2}-2Q_{U_l}I_{3U_l}\right)\right.\nonumber\\&&
+\frac{2+\left(1+g5s\right)\lambda_{HV}}{4s_w^2}\sum_{g}\left(V_{U_l,D_g}V^{\dagger}_{D_g,U_l}\right)
+\frac{1}{2m_W^2s_w^2}\nonumber\\&&
\left.\left(\frac{1}{2}\sum_{g}\left(V_{U_l,D_g}V^{\dagger}_{D_g,U_l}m_{D_g}^2\right)+m_{U_l}^2\left(\frac{1}{4}
+I_{3U_l}^2\right)\right)\right],\nonumber\\
C^{D_l\bar{D}_m}_{-}&=&\delta_{lm}~\frac{1}{16}\left[\left(1+\lambda_{HV}\right)\frac{Q_{D_l}^2}{c_w^2}
+\frac{m_{D_l}^2}{2s_w^2m_W^2}\left(\frac{1}{2}\sum_{g}\left(V_{U_g,D_l}V^{\dagger}_{D_l,U_g}\right)\right.\right.\nonumber\\&&
\left.\left.+\frac{1}{4}+I_{3D_l}^2\right)\right],\nonumber\\
C^{D_l\bar{D}_m}_{+}&=&\delta_{lm}~\frac{1}{16}\left[\left(1+\lambda_{HV}\right)\frac{Q_{D_l}^2}{c_w^2}+\frac{2+
\left(1+g5s\right)\lambda_{HV}}{2c_w^2}\left(\frac{I_{3D_l}^2}{s_w^2}-2Q_{D_l}I_{3D_l}\right)\right.\nonumber\\&&
+\frac{2+\left(1+g5s\right)\lambda_{HV}}{4s_w^2}\sum_{g}\left(V_{U_g,D_l}V^{\dagger}_{D_l,U_g}\right)
+\frac{1}{2m_W^2s_w^2}\nonumber\\&&
\left.\left(\frac{1}{2}\sum_{g}\left(V_{U_g,D_l}V^{\dagger}_{D_l,U_g}m_{U_g}^2\right)+m_{D_l}^2\left(\frac{1}{4}
+I_{3D_l}^2\right)\right)\right].
\end{eqnarray}
\end{center}
\vspace{0.5cm}
\subsubsection{Electroweak effective vertices with 4 external legs}

\begin{center}
\begin{figure}
\hspace{0cm}\includegraphics[width=9cm]{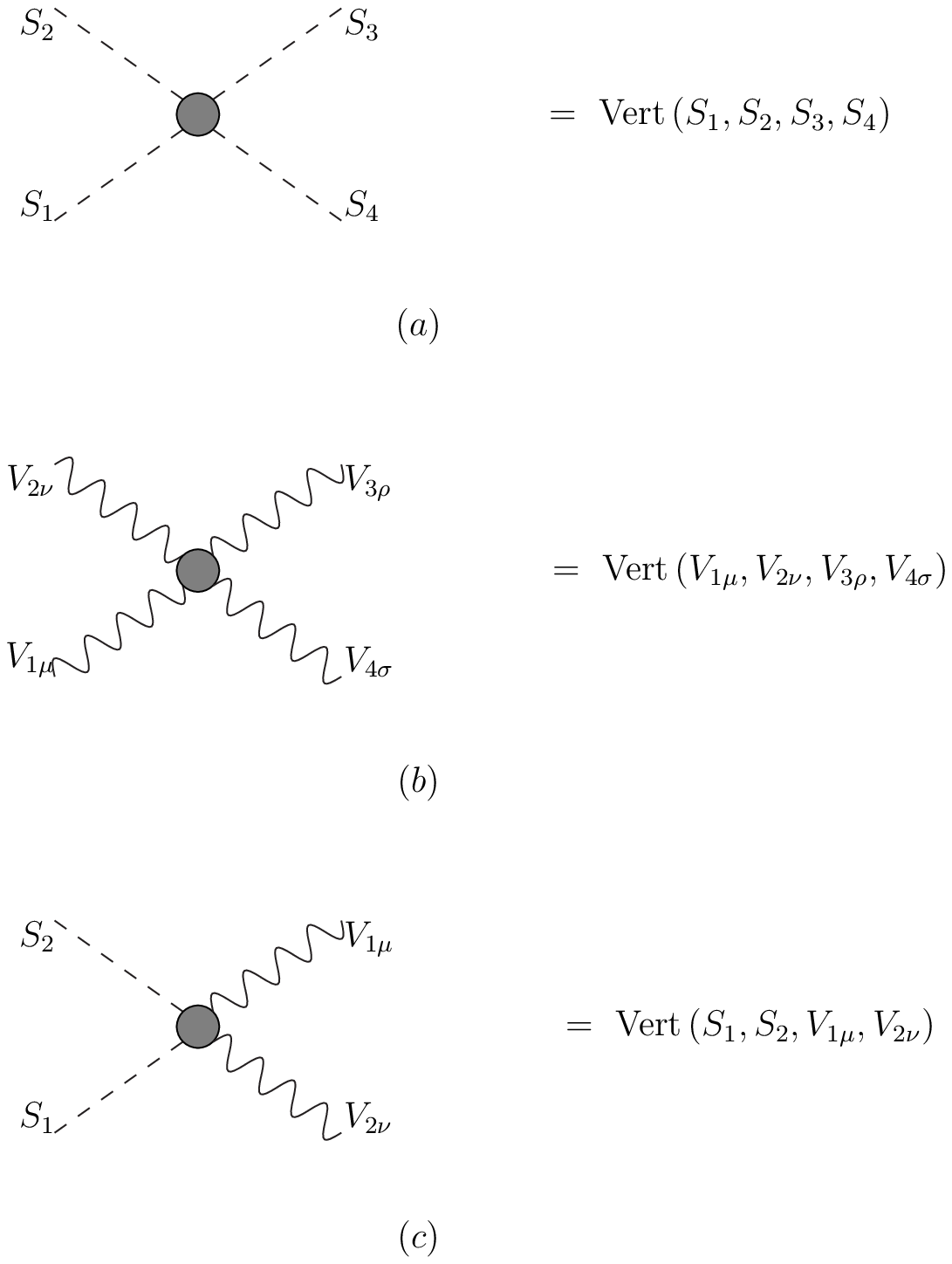}
\caption{\label{fig:ew4} All possible non-vanishing 4-point vertices
in EW.}
\end{figure}
\end{center}

In this part, we list all possible non-vanishing 4-point vertices in
EW, which are shown in Fig.\ref{fig:ew4}.

\begin{center}
\textbf{{\small Scalar-Scalar-Scalar-Scalar vertices}}
 \vspace{0.1cm}
\end{center}

The generic effective vertex is shown in Fig.\ref{fig:ew4}~$(a)$
with the following expression
\begin{center}
\begin{eqnarray}
{\rm Vert}\left(S_1,S_2,S_3,S_4\right)&=& \frac{i e^4}{
\pi^2}~C^{S_1S_2S_3S_4},
\end{eqnarray}
\end{center}
with the actual values of $S_1$, $S_2$, $S_3$, $S_4$ and
$C^{S_1S_2S_3S_4}$
\begin{center}
\begin{eqnarray*}
C^{HHH\phi^0}&=&C^{H\phi^0\phi^0\phi^0}=C^{H\phi^0\phi^-\phi^+}=0,\nonumber\\
C^{HHHH}&=&C^{\phi^0\phi^0\phi^0\phi^0}=\frac{1}{64s_w^4}K_1,\nonumber\\
C^{HH\phi^0\phi^0}&=&\frac{1}{192s_w^4}\left(K_1-\frac{1-g5s}{2}~
\frac{4}{m_W^4}\left(\sum_{l}m_{e_l}^4+N_c\sum_{l}m_{Q_l}^4\right)\right),\nonumber\\
C^{HH\phi^-\phi^+}&=&\frac{1}{64s_w^4}\left\{K_2-\frac{1-g5s}{2}~\frac{2N_c}{3m_W^4}
\right.\nonumber\\&&\left.\sum_{l,m}\left[V_{U_l,D_m}V^{\dagger}_{D_m,U_l}m_{U_l}m_{D_m}
\left(m_{U_l}^2+m_{U_l}m_{D_m}+m_{D_m}^2\right)\right]\right\},
\end{eqnarray*}
\end{center}
\begin{center}
\begin{eqnarray}
C^{\phi^0\phi^0\phi^-\phi^+}&=&\frac{1}{64s_w^4}\left\{K_2+\frac{1-g5s}{2}~\frac{2N_c}{3m_W^4}
\right.\nonumber\\&&\left.\sum_{l,m}\left[V_{U_l,D_m}V^{\dagger}_{D_m,U_l}m_{U_l}m_{D_m}
\left(m_{U_l}^2-m_{U_l}m_{D_m}+m_{D_m}^2\right)\right]\right\},\nonumber\\
C^{\phi^-\phi^+\phi^-\phi^+}&=&\frac{1}{32s_w^4}K_3,
\end{eqnarray}
\end{center}
where
\begin{center}
\begin{eqnarray}
K_1&=&\frac{1}{m_W^2}\left[\frac{5}{m_W^2}\left(\sum_{g}m_{e_l}^4+N_c\sum_{g}m_{Q_l}^4\right)+\frac{3m_H^2}{2}
\left(1+\frac{1}{2c_w^2}\right)\right]\nonumber\\&&+\frac{1-12\lambda_{HV}}{2}\left(1+\frac{1}{2c_w^4}\right),\nonumber\\
K_2&=&\frac{1}{m_W^2}\left[\frac{9+g5s}{6m_W^2}\left(\sum_{l}m_{e_l}^4+N_c\sum_{l,m}V_{U_l,D_m}
V^{\dagger}_{D_m,U_l}\left(m_{U_l}^4+m_{D_m}^4\right)\right)\right.\nonumber\\&&
\left.+\frac{m_H^2}{2}\left(1+\frac{1}{2c_w^2}\right)\right]+\frac{1-12\lambda_{HV}}{4}
\left(1+\frac{s_w^2}{3c_w^2}\left(1+\frac{1}{c_w^2}\right)\right),\nonumber\\
K_3&=&\frac{1}{m_W^2}\left[\frac{9+g5s}{6m_W^2}\left(\sum_{l}m_{e_l}^4+N_c\sum_{i,j,k,l}\left(V_{U_i,D_j}
V^{\dagger}_{D_j,U_k}V_{U_k,D_l}V^{\dagger}_{D_l,U_i}\right.\right.\right.\nonumber\\&&
\left.\left.\left(m_{U_i}^2m_{U_k}^2+m_{D_j}^2m_{D_l}^2+\frac{1-g5s}{4}m_{U_i}m_{U_k}m_{D_j}m_{D_l}
\right)\right)\right)\nonumber\\&&\left.+\frac{m_H^2}{2}\left(1+\frac{1}{2c_w^2}\right)\right]
+\left(1-12\lambda_{HV}\right)\left(\frac{1+s_w^4}{4}+\frac{1}{6}\left(s_w^2+\frac{2s_w^6}{c_w^2}\right)
+\frac{s_w^8}{12c_w^4}\right).
\end{eqnarray}
\end{center}
\vspace{0.5cm}

\begin{center}
\textbf{{\small Vector-Vector-Vector-Vector vertices}}
 \vspace{0.1cm}
\end{center}

The generic effective vertex is shown in Fig.\ref{fig:ew4}~$(b)$
with the following expression
\begin{center}
\begin{eqnarray}
{\rm Vert}\left(V_{1\mu},V_{2\nu},V_{3\rho},V_{4\sigma}\right)&=&
\frac{i e^4}{
\pi^2}~\left(C^{V_1V_2V_3V_4}_1g_{\mu\nu}g_{\rho\sigma}+C^{V_1V_2V_3V_4}_2g_{\mu\rho}g_{\nu\sigma}
\right.\nonumber\\&&\left.+C^{V_1V_2V_3V_4}_3g_{\mu\sigma}g_{\nu\rho}+C^{V_1V_2V_3V_4}_0
\varepsilon_{\mu\nu\rho\sigma}\right),
\end{eqnarray}
\end{center}
with the actual values of $V_1$, $V_2$, $V_3$, $V_4$,
$C_1^{V_1V_2V_3V_4}$,$C_2^{V_1V_2V_3V_4}$, $C_3^{V_1V_2V_3V_4}$ and
$C^{V_1V_2V_3V_4}_0$
\begin{center}
\begin{eqnarray*}
C_1^{AAAA}&=&\frac{1}{12}\left(-1+\sum_{l}Q_{L_l}^4+N_c\sum_{l}Q_{Q_l}^4\right),\nonumber\\
C_2^{AAAA}&=&C_3^{AAAA}=C_1^{AAAA},\nonumber\\
C_0^{AAAA}&=&0,\nonumber\\
C_1^{AAAZ}&=&\frac{1}{12}\left[\frac{c_w}{s_w}+\sum_{l}\left(\frac{s_w}{c_w}Q_{L_l}^4
-\frac{1}{2s_wc_w}Q_{L_l}^3I_{3L_l}\right)\right.\nonumber\\&&
\left.+N_c\sum_{l}\left(\frac{s_w}{c_w}Q_{Q_l}^4-\frac{1}{2s_wc_w}Q_{Q_l}^3I_{3Q_l}\right)\right],\nonumber\\
C_2^{AAAZ}&=&C_3^{AAAZ}=C_1^{AAAZ},\nonumber\\
C_0^{AAAZ}&=&0,\nonumber\\
C^{AAZZ}_1&=&\frac{1}{12}\left[-\frac{c_w^2}{s_w^2}+\frac{1}{2}\sum_{l}\left(\frac{s_w^2}{c_w^2}
Q_{L_l}^4+\left(\frac{s_w}{c_w}Q_{L_l}^2-\frac{1}{s_wc_w}Q_{L_l}I_{3L_l}\right)^2\right)\right.\nonumber\\&&
+\frac{N_c}{2}\sum_{l}\left(\frac{s_w^2}{c_w^2}Q_{Q_l}^4+\left(\frac{s_w}{c_w}Q_{Q_l}^2-\frac{1}{s_wc_w}
Q_{Q_l}I_{3Q_l}\right)^2\right)\nonumber\\&&
\left.+\frac{1-g5s}{2}~2\left(\sum_{l}\frac{1}{s_w^2c_w^2}Q_{L_l}^2I_{3L_l}^2
+N_c\sum_{l}\frac{1}{s_w^2c_w^2}Q_{Q_l}^2I_{3Q_l}^2\right)\right],\nonumber\\
C_2^{AAZZ}&=&\frac{1}{12}\left[-\frac{c_w^2}{s_w^2}+\frac{1}{2}\sum_{l}\left(\frac{s_w^2}{c_w^2}
Q_{L_l}^4+\left(\frac{s_w}{c_w}Q_{L_l}^2-\frac{1}{s_wc_w}Q_{L_l}I_{3L_l}\right)^2\right)\right.\nonumber\\&&
\left.+\frac{N_c}{2}\sum_{l}\left(\frac{s_w^2}{c_w^2}Q_{Q_l}^4+\left(\frac{s_w}{c_w}Q_{Q_l}^2-\frac{1}{s_wc_w}
Q_{Q_l}I_{3Q_l}\right)^2\right)\right],\nonumber\\
C_3^{AAZZ}&=&C_2^{AAZZ},\nonumber\\
C_0^{AAZZ}&=&0,\nonumber\\
C^{AZZZ}_1&=&\frac{1}{12}\left[\frac{c_w^3}{s_w^3}+\sum_{l}\left(\frac{s_w^3}{c_w^3}Q_{L_l}^4
-\frac{3s_w}{2c_w}Q_{L_l}^3I_{3L_l}+\frac{5-2g5s}{2s_wc_w^3}Q_{L_l}^2I_{3L_l}^2\right.\right.\nonumber\\&&
\left.-\frac{2-g5s}{2s_w^3c_w^3}Q_{L_l}I_{3L_l}^3\right)+N_c\sum_{l}
\left(\frac{s_w^3}{c_w^3}Q_{Q_l}^4-\frac{3s_w}{2c_w^3}Q_{Q_L}^3I_{3Q_l}\right.\nonumber\\&&
\left.\left.+\frac{5-2g5s}{2s_wc_w^3}Q_{Q_l}^2I_{3Q_l}^2-\frac{2-g5s}{2s_w^3c_w^3}Q_{Q_l}I_{Q_l}^3\right)\right],\nonumber\\
C^{AZZZ}_2&=&C^{AZZZ}_3=C^{AZZZ}_1,\nonumber\\
C_0^{AZZZ}&=&0,
\end{eqnarray*}
\end{center}
\begin{center}
\begin{eqnarray*}
C^{ZZZZ}_1&=&\frac{1}{12}\left[-\frac{c_w^4}{s_w^4}+\sum_{l}\left(\frac{s_w^4}{c_w^4}Q_{L_l}^4-\frac{2s_w^2}{c_w^2}
Q_{L_l}^3I_{3L_l}+\frac{5-2g5s}{c_w^4}Q_{L_l}^2I_{3L_l}^2\right.\right.\nonumber\\&&
\left.-\frac{4-2g5s}{s_w^2c_w^4}Q_{L_l}I_{3L_l}^3+\frac{5-3g5s}{4s_w^4c_w^4}I_{3L_l}^4\right)\nonumber\\&&
+N_c\sum_{l}\left(\frac{s_w^4}{c_w^4}Q_{Q_l}^4-\frac{2s_w^2}{c_w^4}Q_{Q_l}^3I_{3Q_l}
+\frac{5-2g5s}{c_w^4}Q_{Q_l}^2I_{3Q_l}^2\right.\nonumber\\&&\left.\left.
-\frac{4-2g5s}{s_w^2c_w^4}Q_{Q_l}I_{3Q_l}^3+\frac{5-3g5s}{4s_w^4c_w^4}I_{3Q_l}^4\right)\right],\nonumber\\
C^{ZZZZ}_2&=&C_3^{ZZZZ},\nonumber\\
C_0^{ZZZZ}&=&0,\nonumber\\
C^{AAW^-W^+}_1&=&\frac{1}{16s_w^2}\left[\frac{10+4\lambda_{HV}}{3}+3~\frac{7-g5s}{6}+\frac{\left(51-g5s\right)N_c}{54}\sum_{l,m}
V_{U_l,D_m}V^{\dagger}_{D_m,U_l}\right],\nonumber\\
C^{AAW^-W^+}_2&=&-\frac{1}{16s_w^2}\left[\frac{7+2\lambda_{HV}}{3}+1+\frac{11N_c}{27}\sum_{l,m}V_{U_l,D_m}
V^{\dagger}_{D_m,U_l}\right],\nonumber\\
C^{AAW^-W^+}_3&=&C^{AAW^-W^+}_2,\nonumber\\
C_0^{AAW^-W^+}&=&\frac{1+g5s}{2}~\frac{i}{48s_w^2}\left(N_c\sum_{l,m}\left(V_{U_l,D_m}V^{\dagger}_{D_m,U_l}\right)-9\right),\nonumber\\
C^{AZW^-W^+}_1&=&\frac{1}{16s_wc_w}\left[-\frac{\left(10+4\lambda_{HV}\right)c_w^2}{3s_w^2}
+3\left(\frac{7-g5s}{6}-\frac{10+g5s}{12s_w^2}\right)\right.\nonumber\\&&
\left.+N_c\sum_{l,m}\left(V_{U_l,D_m}V^{\dagger}_{D_m,U_l}\left(\frac{51-g5s}{54}-\frac{10+g5s}{12s_w^2}\right)
V_{U_l,D_m}V^{\dagger}_{D_m,U_l}\right)\right],\nonumber\\
C^{AZW^-W^+}_2&=&\frac{1}{16s_wc_w}\left[\frac{\left(7+2\lambda_{HV}\right)c_w^2}{3s_w^2}
+3\left(\frac{4+g5s}{12s_w^2}-\frac{1}{3}\right)\right.\nonumber\\&&
\left.+N_c\sum_{l,m}
\left(V_{U_l,D_m}V^{\dagger}_{D_m,U_l}\left(\frac{4+g5s}{12s_w^2}-\frac{11}{27}\right)\right)\right],\nonumber\\
C^{AZW^-W^+}_3&=&C^{AZW^-W^+}_2,\nonumber\\
C^{AZW^-W^+}_0&=&\frac{i}{192c_ws_w^3}\left(\frac{1+g5s}{2}~\left(4s_w^2-3\right)+\frac{1-g5s}{2}~2\right)\left(N_c\sum_{l,m}\left(V_{U_l,D_m}V^{\dagger}_{D_m,U_l}\right)-9\right),\nonumber\\
C^{ZZW^-W^+}_1&=&\frac{\left(5+2\lambda_{HV}\right)c_w^2}{24s_w^4}+\frac{1}{16c_w^2}\left[3\left(\frac{7-g5s}{6}
-\frac{10+g5s}{6s_w^2}+\frac{23-g5s}{24s_w^4}\right)\right.\nonumber\\&&
\left.+N_c\left(\frac{51-g5s}{54}-\frac{10+g5s}{6s_w^2}+\frac{23-g5s}{24s_w^4}\right)
\sum_{l,m}\left(V_{U_l,D_m}V^{\dagger}_{D_m,U_l}\right)\right],
\end{eqnarray*}
\end{center}
\begin{center}
\begin{eqnarray}
C^{ZZW^-W^+}_2&=&-\frac{\left(7+2\lambda_{HV}\right)c_w^2}{48s_w^4}+\frac{1}{16c_w^2}\left[
3\left(-\frac{1}{3}+\frac{4+g5s}{6s_w^2}-\frac{9+g5s}{24s_w^4}\right)\right.\nonumber\\&&
\left.+N_c\left(-\frac{11}{27}+\frac{4+g5s}{6s_w^2}-\frac{9+g5s}{24s_w^4}\right)\sum_{l,m}V_{U_l,D_m}
V^{\dagger}_{D_m,U_l}\right],\nonumber\\
C^{ZZW^-W^+}_3&=&C^{ZZW^-W^+}_2,\nonumber\\
C^{ZZW^-W^+}_0&=&-\frac{i\left(12\left(1+g5s\right)+\left(g5s-1\right)c_w^2\right)}{1152s_w^2c_w^2}\left(N_c\sum_{l,m}\left(V_{U_l,D_m}V^{\dagger}_{D_m,U_l}\right)-9\right),\nonumber\\
C^{W^-W^+W^-W^+}_1&=&\frac{1}{16s_w^4}\left[\frac{3+2\lambda_{HV}}{3}+3\frac{7-g5s}{12}
\right.\nonumber\\&&
\left.+\frac{\left(7-g5s\right)N_c}{12}\sum_{i,j,k,m}\left(V_{U_i,D_j}V^{\dagger}_{D_j,U_k}V_{U_k,D_m}V^{\dagger}_{D_m,U_i}
\right)\right],\nonumber\\
C^{W^-W^+W^-W^+}_2&=&-\frac{1}{8s_w^4}\left[\frac{7+2\lambda_{HV}}{3}+3\frac{9+g5s}{24}
\right.\nonumber\\&&\left.+\frac{\left(9+g5s\right)N_c}{24}
\sum_{i,j,k,m}\left(V_{U_i,D_j}V^{\dagger}_{D_j,U_k}V_{U_k,D_m}V^{\dagger}_{D_m,U_i}
\right)\right],\nonumber\\
C^{W^-W^+W^-W^+}_3&=&C^{W^-W^+W^-W^+}_1,\nonumber\\
C_0^{W^-W^+W^-W^+}&=&0.
\end{eqnarray}
\end{center}
\vspace{0.5cm}

\begin{center}
\textbf{{\small Scalar-Scalar-Vector-Vector vertices}}
 \vspace{0.1cm}
\end{center}

The generic effective vertex is shown in Fig.\ref{fig:ew4}~$(c)$
with the following expression
\begin{center}
\begin{eqnarray}
{\rm Vert}\left(S_1,S_2,V_{1\mu},V_{2\nu}\right)&=& \frac{i e^4}{
\pi^2}~C^{S_1S_2V_1V_2}~g_{\mu\nu},
\end{eqnarray}
\end{center}
with the actual values of $S_1$, $S_2$, $V_1$, $V_2$ and
$C^{S_1S_2V_1V_2}$
\begin{center}
\begin{eqnarray*}
C^{H\phi^0AA}&=&C^{H\phi^0AZ}=C^{H\phi^0ZZ}=C^{H\phi^0W^+W^-}=0,\nonumber\\
C^{HHAA}&=&\frac{1}{16s_w^2}\left[\frac{1}{12}-\frac{1}{m_W^2}\left(\sum_{l}\left(Q_{L_l}^2m_{L_l}^2\right)
+N_c\sum_{l}\left(Q_{Q_l}^2m_{Q_l}^2\right)\right)\right],\nonumber\\
C^{\phi^0\phi^0AA}&=&\frac{1}{16s_w^2}\left[\frac{1}{12}-\frac{7-4~g5s}{3m_W^2}\left(\sum_{l}\left(Q_{L_l}^2m_{L_l}^2\right)
+N_c\sum_{l}\left(Q_{Q_l}^2m_{Q_l}^2\right)\right)\right],\nonumber\\
C^{HHAZ}&=&\frac{1}{16s_w}\left[\frac{4+s_w^2}{12s_w^2c_w}+\frac{1}{m_W^2c_w}\left(
\sum_{l}\left(Q_{L_l}m_{L_l}^2\left(\frac{I_{3L_l}}{2s_w^2}-Q_{L_l}\right)\right)\right.\right.\nonumber\\&&
\left.\left.+N_c\sum_{l}\left(Q_{Q_l}m_{Q_l}^2\left(\frac{I_{3Q_l}}{2s_w^2}-Q_{Q_l}\right)\right)\right)\right],\nonumber\\
C^{\phi^0\phi^0AZ}&=&\frac{1}{16s_w}\left[\frac{4+s_w^2}{12s_w^2c_w}+\frac{7-4g5s}{3m_W^2c_w}\left(
\sum_{l}\left(Q_{L_l}m_{L_l}^2\left(\frac{I_{3L_l}}{2s_w^2}-Q_{L_l}\right)\right)\right.\right.\nonumber\\&&
\left.\left.+N_c\sum_{l}\left(Q_{Q_l}m_{Q_l}^2\left(\frac{I_{3Q_l}}{2s_w^2}-Q_{Q_l}\right)\right)\right)\right],\nonumber\\
C^{HHZZ}&=&-\frac{1}{16c_w^2}\left[\frac{1+2c_w^2+40c_w^4-4c_w^6}{48s_w^4c_w^2}\right.\nonumber\\&&
\left.+\frac{1}{m_W^2}
\left(\sum_{l}\left(m_{L_l}^2\left(Q_{L_l}^2+\frac{\left(7+g5s\right)I_{3L_l}^2}{6s_w^4}
-\frac{Q_{L_l}I_{3L_l}}{s_w^2}\right)\right)\right.\right.\nonumber\\&&\left.\left.+N_c\sum_{l}\left(m_{Q_l}^2\left(Q_{Q_l}^2
+\frac{\left(7+g5s\right)I_{3Q_l}^2}{6s_w^4}-\frac{Q_{Q_l}I_{3Q_l}}{s_w^2}\right)\right)\right)\right],\nonumber\\
C^{\phi^0\phi^0ZZ}&=&-\frac{1}{16c_w^2}\left[\frac{1+2c_w^2+40c_w^4-4c_w^6}{48s_w^4c_w^2}\right.\nonumber\\&&
\left.+\frac{7-4g5s}{3m_W^2}
\left(\sum_{l}\left(m_{L_l}^2\left(Q_{L_l}^2+\frac{\left(71+17~g5s\right)I_{3L_l}^2}{66s_w^4}
-\frac{Q_{L_l}I_{3L_l}}{s_w^2}\right)\right)\right.\right.\nonumber\\&&\left.\left.+N_c\sum_{l}\left(m_{Q_l}^2\left(Q_{Q_l}^2
+\frac{\left(71+17~g5s\right)I_{3Q_l}^2}{66s_w^4}-\frac{Q_{Q_l}I_{3Q_l}}{s_w^2}\right)\right)\right)\right],\nonumber\\
C^{HHW^-W^+}&=&-\frac{1}{48s_w^4}\left[\frac{1+38c_w^2}{16c_w^2}\right.\nonumber\\&&\left.+\frac{7+g5s}{8m_W^2}\left(\sum_{l}m_{e_l}^2
+N_c\sum_{l,m}\left(V_{U_l,D_m}V^{\dagger}_{D_m,U_l}\left(m_{U_l}^2+m_{D_m}^2\right)\right)\right)\right],\nonumber\\
\end{eqnarray*}
\end{center}
\begin{center}
\begin{eqnarray*}
C^{\phi^0\phi^0W^-W^+}&=&-\frac{1}{48s_w^4}\left[\frac{1+38c_w^2}{16c_w^2}+\frac{11-3~g5s}{8m_W^2}\left(\sum_{l}m_{e_l}^2
\right.\right.\nonumber\\&&\left.\left.+N_c\sum_{l,m}\left(V_{U_l,D_m}V^{\dagger}_{D_m,U_l}
\left(m_{U_l}^2+m_{D_m}^2-\frac{1-g5s}{2}~\frac{4}{7}m_{U_l}m_{D_m}\right)\right)\right)\right],\nonumber\\
C^{H\phi^+AW^-}&=&C^{H\phi^-AW^+}=\frac{1}{24s_w^3}\left(\frac{1+22c_w^2}{32c_w^2}+K_1\right),\nonumber\\
C^{\phi^0\phi^+AW^-}&=&-\frac{i}{24s_w^3}\left(\frac{1+22c_w^2}{32c_w^2}+K_2\right),\nonumber\\
C^{\phi^0\phi^-AW^+}&=&-C^{\phi^0\phi^+AW^-},\nonumber\\
C^{H\phi^+ZW^-}&=&C^{H\phi^-ZW^+}=\frac{s_w}{c_w}~C^{H\phi^+AW^-},\nonumber\\
C^{\phi^0\phi^+ZW^-}&=&-\frac{i}{24s_w^2c_w}\left(\frac{1+22c_w^2}{32c_w^2}+K_3\right),\nonumber\\
C^{\phi^0\phi^-ZW^+}&=&-C^{\phi^0\phi^+ZW^-},\nonumber\\
C^{\phi^-\phi^+AA}&=&-\frac{1}{12s_w^2}\left[\frac{1+21c_w^2}{16c_w^2}+\frac{1}{m_W^2}\left(\frac{9-g5s}{8}
\sum_{l}m_{e_l}^2\right.\right.\nonumber\\&&
\left.\left.+\frac{5N_c}{6}\sum_{l,m}\left(V_{U_l,D_m}V^{\dagger}_{D_m,U_l}\left(m_{U_l}^2+m_{D_m}^2
+\frac{1-g5s}{2}~\frac{\left(m_{U_l}+m_{D_m}\right)^2}{30}\right)\right)\right)\right],\nonumber\\
C^{\phi^-\phi^+AZ}&=&\frac{1}{12s_wc_w}\left\{\frac{42c_w^4-10c_w^2-1}{32s_w^2c_w^2}\right.\nonumber\\&&
-\frac{1}{m_W^2}\left[
\sum_{l}\left(m_{e_l}^2Q_{e_l}\left(\frac{\left(9-g5s\right)Q_{e_l}}{8}
+\frac{5I_{3\nu_l}}{8s_w^2}\right)\right)\right.\nonumber\\&&
+N_c\sum_{l,m}\left[V_{U_l,D_m}V_{D_m,U_l}^{\dagger}
\left(m_{U_l}^2\left(\frac{5}{6}-\frac{I_{3D_m}}{s_w^2}\left(Q_{D_m}-\frac{5}{8}Q_{U_l}\right)\right)\right.\right.
\nonumber\\&&+m_{D_m}^2\left(\frac{5}{6}-
\frac{I_{3U_l}}{s_w^2}\left(Q_{U_l}-\frac{5}{8}Q_{D_m}\right)\right)
\nonumber\\&&\left.\left.\left.\left.+\frac{1-g5s}{2}~\frac{1}{36}\left(m_{U_l}+m_{D_m}\right)^2
\right)\right]\right]\right\},
\end{eqnarray*}
\end{center}
\begin{center}
\begin{eqnarray}
C^{\phi^-\phi^+ZZ}&=&\frac{1}{12c_w^2}\left\{\frac{-1+2c_w^2+44c_w^4-84c_w^6}{64s_w^4c_w^2}-\frac{1}{m_W^2}
\left[\sum_{l}\left(m_{e_l}^2\left(\frac{\left(9-g5s\right)Q_{e_l}^2}{8}\right.\right.\right.\right.\nonumber\\&&
\left.\left.+\frac{5Q_{e_l}I_{3\nu_l}}{4s_w^2}+\frac{I_{3\nu_l}^2}{s_w^4}\right)\right)
+N_c\sum_{l,m}\left[V_{U_l,D_m}V^{\dagger}_{D_m,U_l}\right.\nonumber\\&&\left(m_{U_l}^2\left(\frac{5}{6}
-\frac{I_{3D_m}}{s_w^2}\left(2Q_{D_m}-\frac{5}{4}Q_{U_l}\right)+\frac{I_{3D_m}^2}{s_w^4}\right)\right.\nonumber\\&&
+m_{D_m}^2\left(\frac{5}{6}
-\frac{I_{3U_l}}{s_w^2}\left(2Q_{U_l}-\frac{5}{4}Q_{D_m}\right)+\frac{I_{3U_l}^2}{s_w^4}\right)\nonumber\\&&
\left.\left.\left.\left.+\frac{1-g5s}{2}~\left(\frac{\left(m_{U_l}+m_{D_m}\right)^2}{36}
+\frac{m_{U_l}m_{D_m}}{8s_w^4}\right)\right)\right]\right]\right\},\nonumber\\
C^{\phi^-\phi^+W^-W^+}&=&-\frac{1}{48s_w^4}\left\{\frac{38c_w^2+1}{16c_w^2}
+\frac{1}{m_W^2}\left[\left(\frac{9-g5s}{8}~\sum_{l}m_{e_l}^2\right.\right.\right.\nonumber\\&&+\frac{N_c}{4}\sum_{i,j,k,l}
\left(V_{U_i,D_j}V^{\dagger}_{D_j,U_k}V_{U_k,D_l}V^{\dagger}_{D_l,U_i}
\left(2m_{U_l}^2+2m_{U_k}^2+2m_{D_j}^2+2m_{D_l}^2\right.\right.\nonumber\\&&
+\frac{1-g5s}{2}~\left(m_{U_i}m_{U_k}+m_{U_i}m_{D_j}+m_{U_i}m_{D_l}+m_{U_k}m_{D_j}\right.\nonumber\\&&
\left.\left.\left.\left.\left.\left.+m_{U_k}m_{D_l}
+m_{D_j}m_{D_l}\right)\right)\right)\right)\right]\right\}.
\end{eqnarray}
\end{center}
where
\begin{center}
\begin{eqnarray}
K_1&=&\frac{1}{8m_W^2}\left[\sum_{l}m_{e_l}^2+N_c\sum_{l,m}\left(V_{U_l,D_m}V^{\dagger}_{D_m,U_l}
\left(3m_{D_m}^2+2m_{U_l}^2\right)\right)\right],\nonumber\\
K_2&=&\frac{1}{8m_W^2}\left[\left(-1+2g5s\right)\sum_{l}m_{e_l}^2\right.\nonumber\\&&
\left.+N_c\sum_{l,m}\left(V_{U_l,D_m}V^{\dagger}_{D_m,U_l}
\left(\frac{7+2g5s}{3}m_{D_m}^2+\frac{2+4g5s}{3}m_{U_l}^2\right)\right)\right],\nonumber\\
K_3&=&\frac{1}{8m_W^2}\left[\left(-1+2g5s\right)\sum_{l}m_{e_l}^2+\frac{1-g5s}{2}
~\frac{1}{s_w^2}\sum_{l}m_{e_l}^2\right.\nonumber\\&&
+N_c\sum_{l,m}\left(V_{U_l,D_m}V^{\dagger}_{D_m,U_l}
\left(\frac{7+2g5s}{3}m_{D_m}^2+\frac{2+4g5s}{3}m_{U_l}^2\right.\right.\nonumber\\&&
\left.\left.\left.+\frac{1-g5s}{2}~\frac{m_{U_l}^2+m_{D_m}^2}{s_w^2}\right)\right)\right].
\end{eqnarray}
\end{center}
 \vspace{0.5cm}

All possible non-vanishing $R$ effective vertices in SM are
 listed above. Our results are the same as those in
Refs.\cite{Draggiotis:2009yb,Garzelli:2009is} in the same $\gamma_5$
scheme.  In addition, all terms proportional to $\varepsilon$
evidently vanish in the formulae after including all fermions in SM.
This is guaranteed by the cancelation of gauge anomaly in SM.
\section{Summary \label{sec:5}}
In summary, we have studied Feynman rules for the rational part $R$
of the Standard Model amplitudes at one-loop level in the 't
Hooft-Veltman $\gamma_5$ scheme. Comparing Feynman rules obtained in
this scheme for quantum chromodynamics and electroweak 1-loop
amplitudes with that obtained in another $\gamma_5$ scheme (the KKS
scheme) in Refs.\cite{Draggiotis:2009yb,Garzelli:2009is}, we find
the latter result can be recovered after setting $g5s=1$ in our
results. As an independent check, we also calculated Feynman rules
in the KKS scheme, and found results completely in agreement with
the analytical expressions given in their updated version, as
mentioned in Ref.\cite{Garzelli:2010}.

With these rational terms, one can simplify fermion chains or Dirac
traces in 4 dimensions at the amplitude level, resulting in more
convenient general calculations for multi-leg processes, compared
with directly solving $d$ dimension Dirac algebras.  To clarify some
existing ambiguities in DREG, we give our results in FDH and HV
DREG, as well as in KKS and HV $\gamma_5$ schemes. To be more
reliable, our results are checked and found to be in agreement with
previously published works (see
Refs.\cite{Draggiotis:2009yb,Garzelli:2009is} and
Ref.\cite{Garzelli:2010}) in the KKS $\gamma_5$ scheme. In
particular, the expressions for $R$ in the HV $\gamma_5$ scheme are
given in this paper to meet the needs of some specific NLO
calculations.

As the HV $\gamma_5$ scheme is the only rigorously proved consistent
regime to all orders, our results in this $\gamma_5$ scheme are
surely useful and necessary in clarifying the $\gamma_5$ problem in
DREG. The only dimensional regularization dependent terms (except
scaleless bubbles) in one-loop calculations are rational terms $R$.
These Feynman rules are also helpful in eliminating or finding the
uncertainties arising from freedoms in DREG. For instance, in NLO
QCD corrections to the $W$ boson hadronic decays, the rational terms
$R$ in HV and KKS $\gamma_5$ schemes are different (see
Eq.(\ref{eq:wqq})), being right-handed in HV scheme while
left-handed in KKS scheme. These differences are the only
differences after including virtual parts, counter terms, and real
radiations in these two $\gamma_5$ schemes. Actually, the violation
of the anti-commutation relation between $\gamma_5$ and
$\gamma_{\mu}$ in HV spoils the Ward identity, and a finite
renormalization is needed in this scheme to obtain the same
expressions in HV and KKS (see the details in
Ref.\cite{Shao:2011zz}).

\begin{acknowledgments}
We would like to thank R. Pittau for kindly checking some of our
results and helpful communications. We also thank Y.Q. Ma, K. Wang
and D. Li for the assistance in writing this paper. This work was
supported by the National Natural Science Foundation of China
(No.10805002, No.11021092, No.11075002, No.11075011), the Foundation
for the Author of National Excellent Doctoral Dissertation of China
(Grant No. 201020), and the Ministry of Science and Technology of
China (2009CB825200).

\end{acknowledgments}




\providecommand{\href}[2]{#2}\begingroup\raggedright\endgroup

\end{document}